\documentclass[11pt]{article}
\pdfoutput=1
\usepackage{soul}
\usepackage{jheppub}
\usepackage{amsfonts}
\usepackage{amsmath}
\allowdisplaybreaks[4]         
\usepackage{amssymb}   
\usepackage{euscript}       
\usepackage{xcolor}          
\usepackage{tensor}     
\usepackage{caption}
\usepackage{subcaption}   

\newcommand{\req}[1]{(\ref{#1})} 

\newcommand{\bea}{\begin{eqnarray}}
\newcommand{\eea}{\end{eqnarray}}
\newcommand{\ba}{\begin{eqnarray}}
\usepackage{braket}
\newcommand{\ea}{\end{eqnarray}}

\newcommand{\beq}{\begin{equation}}
\newcommand{\eeq}{\end{equation} }
\newcommand{\beqa}{\begin{eqnarray}}

\newcommand{\eeqa}{\end{eqnarray}}
\newcommand{\beqar}{\begin{eqnarray*}}
\newcommand{\eeqar}{\end{eqnarray*}}

\newcommand{\be}{\begin{equation}}
\newcommand{\ee}{\end{equation}}
\newcommand{\diff}{\mathrm{d}}

\newcommand{\dv}{\\\notag &}
\newcommand{\dvv}{\right.\\\notag &\left.}
\newcommand{\dvvv}{\right.\right.\\\notag &\left.\left.}

\newcommand{\dvvtag}{\right.\\ &\left.}
\newcommand{\dvvvtag}{\right.\right.\\&\left.\left.}

\renewcommand{\req}[1]{(\ref{#1})}

\newcommand{\E}{\mathcal{E}}

\title{
The extremal Kerr entropy in higher-derivative gravities

}

\author[a]{Pablo A. Cano,}

\author[a]{Marina David}

\affiliation[a]{Instituut voor Theoretische Fysica, KU Leuven.
Celestijnenlaan 200D, B-3001 Leuven, Belgium \vspace{0.1cm}}

\emailAdd{pabloantonio.cano@kuleuven.be}
\emailAdd{marina.david@kuleuven.be}

\date{\today}
\abstract{
We investigate higher derivative corrections to the extremal Kerr black hole in the context of heterotic string theory with $\alpha'$ corrections and of a cubic-curvature extension of general relativity. 
By analyzing the near-horizon extremal geometry of these black holes, we are able to compute the Iyer-Wald entropy as well as the angular momentum via generalized Komar integrals. 
In the case of the stringy corrections, we obtain the physically relevant relation $S(J)$ at order $\alpha'^2$. On the other hand, the cubic theories, which are chosen as Einsteinian cubic gravity plus a new odd-parity density with analogous features, possess special integrability properties that enable us to obtain exact results in the higher-derivative couplings. This allows us to find the relation $S(J)$ at arbitrary orders in the couplings and even to study it in a non-perturbative way. We also extend our analysis to the case of the extremal Kerr-(A)dS black hole.
}

\begin{document} 
\maketitle
\flushbottom

\newpage
\section{Introduction}
\label{sec:Introduction}

	The investigation of higher-derivative corrections in general relativity (GR) provides us with key insights in our understanding of quantum gravity. In fact, the Einstein-Hilbert Lagrangian is considered to be just the leading term in an effective action which is expected to contain an infinite tower of higher-derivative terms. This is in particular realized in string theory \cite{Gross:1986iv,Gross:1986mw,Bergshoeff:1989de}, although the kind of corrections one can obtain in four dimensions depend greatly on the string type and on the compactification process.
	For this reason, in many instances it is useful to follow an effective field theory approach, which allows one to capture the most general correction to GR at any given order in the derivative expansion once the massless degrees of freedom have been specified.
	
	From a theoretical standpoint, the study of higher-derivative corrections to the thermodynamic properties of black holes, like the entropy, is especially relevant. This is because black hole entropy allows one to connect the effective gravitational action with its quantum-gravitational origin whenever a microscopic description of the system is known, \textit{e.g.}, \cite{Strominger:1996sh,Maldacena:1996gb,Johnson:1996ga,Sen:2005wa,Sen:2005iz,Sen:2007qy,Benini:2015eyy,Zaffaroni:2019dhb}.

    However, computing higher-derivative corrections to black hole geometries, even at low orders in the derivative expansion, can be challenging. 
    In particular, there is still a gap in understanding the solution space for rotating backgrounds. One approach that has been taken in the context of corrections to Kerr black holes is to consider an expansion around slow rotation \cite{Yunes:2009hc,Pani:2011gy,Cardoso:2018ptl,Cano:2019ore}. This allows us to perturbatively solve for the fields of the theory at a given order in the angular momentum. However, the solution in the extremal limit cannot be captured by this approach, and one often has to restrict to numeric methods \cite{Kleihaus:2011tg,Delsate:2018ome} with the corresponding loss of analytic control.
    
    If one is not interested in the full black hole solution but only on its thermodynamic properties, there are several strategies that can be followed. 
    One possibility consists in evaluating the Euclidean action on the uncorrected black hole solution \cite{Reall:2019sah,Melo:2020amq, Bobev:2022bjm,Cassani:2022lrk}. As shown by \cite{Reall:2019sah} the leading-order corrections to black hole thermodynamics can be obtained in this way, hence avoiding the intricate task of solving the equations of motion. 
	In the case of extremal black holes, another approach is to investigate the solutions at a specific region of the spacetime, namely, the near-horizon geometry. What makes this favorable is that the near-horizon metric of extremal black holes has additional symmetries, mainly an enhanced $\mathrm{SO}(2,1)$ that does not exist in the full solution. In the case of rotating black holes in four dimensions, this becomes a $\mathrm{U}(1)$ fibered over the $\mathrm{SO}(2,1)$. This enhancement of isometries is not altered by higher derivative terms  and  therefore we can take advantage of the near-horizon extremal geometries (NHEG) to study corrections to the black hole thermodynamic properties. In addition,  the near-horizon approach  can in principle be applied beyond leading-order corrections, which is one of the main motivations of this paper.

	Investigating the NHEGs has shown to be quite fruitful. In fact, it has been well studied that the entropy of extremal black holes can be computed by what is now known as the entropy function formalism \cite{Sen:2005wa, Sen:2005iz,Dabholkar:2006tb,Sen:2007qy}.\footnote{In the context of AdS/CFT, the near-horizon geometry can also be applied to study the entropy via the dual two-dimensional CFT \cite{Chow:2008dp,David:2020ems,David:2020jhp}, including in the near-extremal limit \cite{Larsen:2020lhg,David:2021qaa}.} The strength of this method precisely comes from the enhancement of isometries at the horizon in the extremal limit. The first examples of the entropy function were for spherically symmetric black holes \cite{Sen:2005wa} and later it was extended to rotating solutions \cite{Astefanesei:2006dd,Morales:2006gm,LopesCardoso:2007hen}.  In the static case, the equations of motion for NHEGs are reduced to a system of algebraic equations for the parameters of the solution, hence making the resolution essentially straightforward. 
	Rotating NHEGs, on the other hand, require to solve  a system of differential equations, which become ever more involved with the inclusion of higher-derivative corrections. Thus, even though the entropy function formalism provides a general framework to study these near horizon geometries, important work has to be carried out on a case-by-case basis. 
	
	On the other hand, in order to obtain a physically meaningful result, the entropy must be expressed in terms of other thermodynamic quantities of the black hole, like its charges. Therefore, the correct identification of the black hole charges from the NHEG is also crucial. In the case of the angular momentum $J$, a first-principle computation involves the evaluation of the Noether charge associated to Killing vector that generates rotations \cite{Lee:1990nz,Wald:1993nt,Iyer:1994ys}. Such charge is defined as the integral of the Noether charge two-form at infinity and cannot be in principle defined on the near-horizon geometry. Here we make use of generalized Komar integrals \cite{Komar:1958wp,Bazanski:1990qd,Kastor:2008xb, Kastor:2009wy, Ortin:2021ade}, which are a modification of the Noether charge independent of the surface of integration, in order to provide a first-principle computation of the angular momentum from the NHEG.  We note that this approach is in principle different to the one used in \cite{Astefanesei:2006dd} to define the angular momentum.\footnote{In fact, one must be careful with the identification of charges via the entropy function due to the existence of different notions of charge (\textit{i.e.}, Page, Maxwell or brane-source charges \cite{Marolf:2000cb}). Interpreting the charges correctly is especially non-trivial in the presence of higher-derivative corrections \cite{Cano:2018hut,Faedo:2019xii,Cano:2021dyy}.}

	In this paper, we apply the methods discussed above to obtain the corrections to the relation $S(J)$ for extremal Kerr black holes in two relevant higher-derivative theories. The first is motivated by the effective action of heterotic string theory upon compactification on a six-torus. The other is a general effective field theory extension of GR with six-derivative terms which we choose in a specific way. In the following, we provide an overview of these theories along with our main results. 
	
\subsection{Summary of results}

\subsubsection*{Heterotic string theory with $\alpha'$ corrections:}

We consider the following string-inspired theory\footnote{Up to the topological Gauss-Bonnet term $\mathcal{X}_{4}$, this theory is known as dynamical Chern-Simons gravity \cite{CAMPBELL1991778,Alexander:2009tp}.}
\begin{equation}\label{Istringy}
I_{\rm stringy}=\frac{1}{16\pi G}\int d^4x\sqrt{|g|}\left[R-\frac{1}{2}(\partial\psi)^2-\frac{\alpha'}{8}\psi\tensor{R}{_{\mu\nu\rho\sigma}}\tensor{\tilde{R}}{^{\mu\nu\rho\sigma}}+\frac{\alpha'}{8}\mathcal{X}_{4} \right]\, ,
\end{equation}
where $\alpha'=\ell_{s}^2$ is the square of the string length scale, 

\begin{equation}\label{dualriem}
\tensor{\tilde{R}}{^{\mu\nu\rho\sigma}}=\frac{1}{2}\epsilon^{\mu\nu\alpha\beta}\tensor{R}{_{\alpha\beta}^{\rho\sigma}}
\end{equation}
is the dual Riemann tensor and 

\begin{equation}\label{GB}
\mathcal{X}_{4}=R^2-4R_{\mu\nu}R^{\mu\nu}+R_{\mu\nu\rho\sigma}R^{\mu\nu\rho\sigma}
\end{equation}
is the Gauss-Bonnet density. We will argue that such theory is obtained from heterotic string theory upon dimensional reduction and stabilization of the dilaton. For this theory we are able to find explicitly the corrections to the near-horizon extremal Kerr (NHEK) metric and we obtain that the entropy of the extremal Kerr black hole is corrected according to\footnote{Note that the entropy is a function of the absolute value $|J|$, but throughout the paper we will assume $J>0$ hence avoiding using absolute values.}

\begin{equation} \label{eq: entropy for heterotic string intro}
	S = 2 \pi  |J|+\alpha'\frac{\pi}{2G}-\alpha'^2 \frac{\pi}{|J| G^2}\left(\frac{493 }{3360}+\frac{3 \pi}{128}\right)+O(\alpha'^{3})\, .
\end{equation}

\subsubsection*{Cubic gravity}
We also consider the following extension of GR,
\begin{equation}\label{Icubic}
\begin{aligned}
I_{\rm cubic}=&\frac{1}{16\pi G}\int d^4x\sqrt{|g|}\left[R+\lambda\mathcal{P}+\tilde\lambda \tilde{\mathcal{P}}\right]\, ,
\end{aligned}
\end{equation}
where $\mathcal{P}$ and $\tilde{\mathcal{P}}$ are the following even- and odd-parity cubic-curvature terms
\begin{align}\label{ECG}
\mathcal{P}&=12 \tensor{R}{_{\mu}^{\rho}_{\nu}^{\sigma}}\tensor{R}{_{\rho}^{\alpha}_{\sigma}^{\beta}}\tensor{R}{_{\alpha}^{\mu}_{\beta}^{\nu}}+\tensor{R}{_{\mu\nu}^{\rho\sigma}}\tensor{R}{_{\rho\sigma}^{\alpha\beta}}\tensor{R}{_{\alpha\beta}^{\mu\nu}}-12R_{\mu\nu\rho\sigma}R^{\mu\rho}R^{\nu\sigma}+8\tensor{R}{_{\mu}^{\nu}}\tensor{R}{_{\nu}^{\rho}}\tensor{R}{_{\rho}^{\mu}}\, ,\\
\tilde{\mathcal{P}}&=7\tensor{R}{^{\mu}^{\nu}_{\alpha}_{\beta}}\tensor{R}{^{\alpha}^{\beta}_{\rho}_{\sigma}}\tensor{\tilde R}{^{\rho}^{\sigma}_{\mu}_{\nu}}-\frac{9}{2}R\tensor{R}{_{\mu\nu\rho\sigma}}\tensor{\tilde{R}}{^{\mu\nu\rho\sigma}}-20\tilde{R}^{\mu\nu\rho\sigma}R_{\mu\rho}R_{\nu\sigma}\, .\label{EvilECG}
\end{align}
On the one hand, up to field redefinitions, this Lagrangian provides a general effective field theory extension of GR to six derivatives \cite{Cano:2019ore,Bueno:2019ltp}. On the other, the cubic operators are chosen in this particular form for a reason: they have special properties that allow us to perform exact (rather than perturbative) computations. In fact, the even-parity Lagrangian \req{ECG} is nothing but Einsteinian cubic gravity (ECG) \cite{Bueno:2016xff}. One of the most interesting aspects about this theory is that the thermodynamic properties of its black hole solutions can be obtained analytically. Indeed, the black hole solutions of ECG have been thoroughly studied \cite{Hennigar:2016gkm,Bueno:2016lrh,Feng:2017tev,Bueno:2018xqc,Bueno:2018uoy,Cano:2019ozf,Frassino:2020zuv,Adair:2020vso}.  
Generalizations of this theory at higher orders and in other dimensions exist \cite{Hennigar:2017ego,Bueno:2017sui,Ahmed:2017jod,Bueno:2017qce,Bueno:2019ycr,Bueno:2022res,Chen:2022fdi}, but so far only parity-preserving generalizations have been studied. 
The density \req{EvilECG} is a new odd-parity density with similar properties to ECG that we are introducing here for the first time. 

For the theory \req{Icubic} we find that the exact relation $S(J)$ can be obtained by solving a system of algebraic equations. In the perturbative limit, this yields

\begin{align}
	S &= 2 \pi  |J|\left[1-\frac{\lambda }{(GJ)^2}+\frac{20 \lambda ^2+9 \tilde{\lambda}^2}{2 (G J)^4}-\frac{184 \lambda^3+117 \lambda  \tilde\lambda^2}{2 (G J)^6}+O\left(J^{-8}\right)\right]\, . \label{eq: S cubic}
\end{align}
The leading correction matches precisely with \cite{Reall:2019sah}, after taking into account that $\lambda=7\eta_{e}$. Our approach allows us to go beyond the first-order correction, and in particular our results show for the first time the effect of parity-violating terms on the entropy, which enter at order $\tilde\lambda^2$.

For both theories, \req{Istringy} and \req{Icubic}, we also extend these results to the case of the extremal Kerr-(A)dS black hole by including a cosmological constant. The resulting expressions are more involved and can be found in \req{eq: S with Lambda zeta heterotic} and \req{eq:ScubicLambda}.

This paper is organized as follows. In section~\ref{sec:charges}, we review the procedure for defining conserved charges in gravity along with setting the conventions and notations used throughout the paper. In section~\ref{sec:heterotic}, we delve into heterotic string theory with higher derivative corrections and compute the corrected near-horizon extremal geometry along with the thermodynamic quantities via the Komar integral and Iyer-Wald entropy. We then move to Einstein gravity with six-derivative corrections in section~\ref{sec:cubic} and investigate how the angular momentum and black hole entropy are affected by the even and odd parity cubic terms. Some final remarks can be found in \ref{sec:conclusions} and additional details of the computation are summarized in appendix~\ref{appendix: heterotic string} and appendix~\ref{appendix: charge2form}.

\section{Conserved charges}
\label{sec:charges}
We review the general formalism of constructing the Noether charges associated to the Killing isometries of the gravitational action.  Let us consider a general higher-curvature theory of the form\footnote{The same results discussed here apply if the theory also contains scalar fields.} 

\begin{equation}
I=\int d^{D}x\sqrt{|g|}\mathcal{L}\left(g^{\mu\nu},R_{\mu\nu\alpha\beta}\right)\, .
\end{equation} 
This theory has equations of motion \cite{Padmanabhan:2011ex}

\begin{equation}\label{Eomgeneral}
\mathcal{E}_{\mu\nu}=\frac{1}{\sqrt{|g|}}\frac{\delta I}{\delta g^{\mu\nu}}=\tensor{R}{_{\mu}^{\sigma\alpha\beta}}P_{\nu\sigma\alpha\beta}-\frac{1}{2}\mathcal{L}g_{\mu\nu}+2\nabla^{\alpha}\nabla^{\beta}P_{\mu\alpha\nu\beta}\, ,
\end{equation}
where 

\begin{equation}\label{defofP1}
P_{\mu\nu\alpha\beta}=\frac{\partial \mathcal{L}}{\partial R^{\mu\nu\alpha\beta}}\, .
\end{equation}
Given the existence of a Killing vector $\xi^{\mu}$, one finds that the $(D-1)$-form Noether current is given by\footnote{We use the convention of denoting differential forms with boldface.} \cite{Wald:1993nt,Iyer:1994ys,Azeyanagi:2009wf,Bueno:2016ypa}
\begin{equation} \label{eq: definition of Noether current}
\boldsymbol{J}_{\xi}=\diff \mathbf{Q}_{\xi}+2\boldsymbol{\epsilon_{\mu}}\xi^{\nu}\tensor{\mathcal{E}}{^{\mu}_{\nu}}\, ,
\end{equation}
where the second term is proportional to the equations of motion and the Noether charge $(D-2)$ form reads\footnote{We define $\boldsymbol{\epsilon_{\mu_1\ldots \mu_{n}}}=\frac{1}{(D-n)!}\epsilon_{\mu_1\ldots\mu_{n}\nu_1\ldots\nu_{D-n}}dx^{\nu_{1}}\wedge\ldots \wedge dx^{\nu_{D-n}}$ and $\epsilon_{0123\dots}=\sqrt{|g|}\epsilon^{0123\dots}$}  
\begin{equation}
	\mathbf{Q}_{\xi}=-\boldsymbol{\epsilon}_{\mu\nu}\left(\bar P^{\mu\nu\alpha\beta} 	\nabla_{\alpha}\xi_{\beta}+2\nabla_{\beta}\bar P^{\mu\nu\alpha\beta}\xi_{\alpha}\right).
\end{equation}
Here, the bar over the tensor $\bar P_{\mu\nu\alpha\beta}$ denotes explicitly that this tensor must satisfy the algebraic Bianchi identity $\bar P_{\mu[\nu\alpha\beta]}=0$, and we can generically write it as 
\begin{equation} \label{eq: def of P}
\bar P_{\mu\nu\alpha\beta}= P_{\mu\nu\alpha\beta}-P_{\mu[\nu\alpha\beta]}\, .
\end{equation}
Note that, depending on how the derivative with respect to the Riemann tensor is defined in \req{defofP1}, the tensor $P_{\mu\nu\alpha\beta}$ may not satisfy the Bianchi identity. It turns out that the antisymmetric part $P_{\mu[\nu\alpha\beta]}$ is irrelevant for the equations of motion \req{Eomgeneral} and for the Iyer-Wald entropy --- see \req{eq: IW entropy formula} below. However, it is crucial that we use $\bar P_{\mu\nu\alpha\beta}$ and not $P_{\mu\nu\alpha\beta}$ in the Noether charge.

Now, on-shell $\E_{\mu\nu}=0$ and the exterior derivative of the current in \eqref{eq: definition of Noether current} yields $\diff \boldsymbol{J}_{\xi}=0$. Then, the total charge associated to $\xi^{\mu}$ can be found by integrating $\boldsymbol{J}_{\xi}$ on any spatial hypersurface of the spacetime, $V_{(D-1)}$. By applying Stokes' theorem, we can then reduce the integration to one over the boundary of $V_{(D-1)}$, which can be typically considered to be the sphere at infinity $\mathbb{S}^{(D-2)}_{\infty}$. The total charge reads
\begin{equation}
Q_{\xi}=\int_{\mathbb{S}^{(D-2)}_{\infty}} \mathbf{Q}_{\xi}\, .
\end{equation}
Note that this charge does not satisfy a Gauss law (\textit{i.e.}, it is not localized), since $\diff  \mathbf{Q}_{\xi}\neq 0$, so it is important that the integration is carried out at infinity. Integrating over any other surface will yield a different result. To remedy this ambiguity, we note that on-shell the Noether current can be written as

\begin{equation}
\boldsymbol{J}_{\xi}=-\frac{1}{2}\star\boldsymbol{\xi}\mathcal{L}\, ,
\end{equation}
where $\star$ is the Hodge dual and $\boldsymbol{\xi}$ is the Killing vector expressed as a one-form.
Then, following \cite{Kastor:2008xb, Ortin:2021ade}, we can define an improved charge $(D-2)$-form (the Noether-Komar charge form) by 


\begin{equation}\label{Komar2form}
\tilde{\mathbf{Q}}_{\xi}=-\boldsymbol{\epsilon}_{\mu\nu}\left(\bar P^{\mu\nu\alpha\beta}\nabla_{\alpha}\xi_{\beta}+2\nabla_{\beta}\bar P^{\mu\nu\alpha\beta}\xi_{\alpha}\right)-\boldsymbol{\Omega}\, ,
\end{equation}
where $\boldsymbol{\Omega}$ is a 2-form satisfying

\begin{equation}\label{omegaformequation}
\diff\boldsymbol{\Omega}=-\frac{1}{2}\star\boldsymbol{\xi}\mathcal{L}\\.
\end{equation}
Note that the local existence of this 2-form is guaranteed by the fact that $d(\star\boldsymbol{\xi}\mathcal{L})=0$ if $\xi^{\nu}$ is a Killing vector. Also,  $\boldsymbol{\Omega}$ is determined up to to a closed form, 
\begin{equation}\label{omegaamb}
\boldsymbol{\Omega}\rightarrow \boldsymbol{\Omega}+\boldsymbol{\beta}\, , \quad \text{with}\quad  \diff\boldsymbol{\beta}=0.
\end{equation}  Now, the Noether-Komar form satisfies a Gauss law,
\begin{equation}
\diff \tilde{\mathbf{Q}}_{\xi}=0\, ,
\end{equation}
and hence the total Komar charge
\begin{equation}
\tilde Q_{\xi}=\int_{\Sigma} \tilde{\mathbf{Q}}_{\xi}
\end{equation}
is independent of the choice of surface of integration $\Sigma$, as long as this is a closed co-dimension two surface homeomorphic to the sphere at infinity. Then, it is not hard to see that this Komar charge is related to the Noether charge of interest by

\begin{equation} \label{eq: Q equals Q plus int of omega}
Q_{\xi}=\tilde Q_{\xi}+\int_{\mathbb{S}^{(D-2)}_{\infty}} \boldsymbol{\Omega}\, .
\end{equation}
Therefore, $\tilde Q_{\xi}$ equals the Noether charge if the integral of $\boldsymbol{\Omega}$ at infinity vanishes. This can always be achieved by using the ambiguity in $\boldsymbol{\Omega}$ in \req{omegaamb}. In particular, one can always add to $\boldsymbol{\Omega}$ a term proportional to the volume element of the integration surface. On the other hand, the addition of an exact form to $\boldsymbol{\Omega}$ does not change the charge. Finally, we must remember that Stokes' theorem applies for regular $p$-forms. Thus, the integral is independent on the surface $\Sigma$ as long as $\tilde{\boldsymbol{Q}}_{\xi}$ is regular. Therefore, when choosing $\boldsymbol{\Omega}$ we must also make sure that this is a regular $(D-2)$-form  everywhere.
Thus, once \emph{any} (regular) $\boldsymbol{\Omega}$ satisfying 
\begin{equation}\label{omegainfty}
\int_{\mathbb{S}^{(D-2)}_{\infty}} \boldsymbol{\Omega}=0 
\end{equation}
has been chosen, the Komar integral computes the Noether charge for any integration surface $\Sigma$.
For example, this allows us to evaluate the charge with $\Sigma$ being the horizon surface. What makes this advantageous is that it is now sufficient to find the near-horizon solution of the black hole of interest, sidestepping the difficulties of analyzing the solution in the full spacetime. However, there is still the question of how one chooses an $\boldsymbol{\Omega}$ satisfying \req{omegainfty} by using only the near-horizon solution. Fortunately, we will see that this can be achieved unambiguously in the case of the angular momentum for stationary and axisymmetric spacetimes. 

In sum, the angular momentum will be given by
\begin{equation} \label{eq: def of J}
J=-2\int_{\Sigma}\left[\boldsymbol{\epsilon}_{\mu\nu}\left(\bar{P}^{\mu\nu\alpha\beta}\nabla_{\alpha}\xi_{\beta}+2\nabla_{\beta}\bar{P}^{\mu\nu\alpha\beta}\xi_{\alpha}\right)+\boldsymbol{\Omega}\right]\, ,\quad \xi=\partial_{\phi}\, ,
\end{equation}
where here, $\Sigma$ is any surface enclosing the black hole horizon, $\partial_{\phi}$ is the Killing vector that generates rotations with periodicity $2\pi$ and we added the appropriate normalization factor  \cite{JKatz_1985}.

Besides the angular momentum, we are interested in computing the entropy of black holes. The prescription to compute the entropy over the horizon $\mathcal{H}$ is given by the Iyer-Wald entropy formula \cite{Wald:1993nt,Iyer:1994ys}
\begin{equation} \label{eq: IW entropy formula}
S=-2\pi \int_{\mathcal{H}}d^{D-2}x\sqrt{h}P^{\mu\nu\alpha\beta}\epsilon_{\mu\nu}\epsilon_{\alpha\beta} ,
\end{equation}
where $P^{\mu\nu\alpha\beta}$ is given by \eqref{defofP1} (or equivalently in this case \eqref{eq: def of P}), $h$ the determinant of the induced $(D-2)$-dimensional horizon metric and $\epsilon_{\mu\nu}$ is the binormal to the horizon, normalized according to $\epsilon_{\mu\nu}\epsilon^{\mu\nu}=-2$.

\section{Heterotic string theory}
\label{sec:heterotic}

\subsection{The four-dimensional stringy action} \label{subsection: the four dimensional heterotic action}

As shown in \cite{Cano:2021rey}, a dimensional reduction of the effective action of heterotic superstring theory on a 6-torus yields the following theory at first order in $\alpha'$
\begin{equation}\label{eq: heterotic action}
I_{\text{stringy}}=\frac{1}{16\pi G}\int d^4x\sqrt{|g|}\left[R-\frac{1}{2}(\partial\varphi)^2-\frac{1}{2}e^{2\varphi}(\partial\psi)^2+\frac{\alpha'}{8}\left(e^{-\varphi}\mathcal{X}_{4}-\psi \tensor{R}{_{\mu\nu\rho\sigma}}\tensor{\tilde{R}}{^{\mu\nu\rho\sigma}} \right)\right]\, .
\end{equation}
Here $\varphi$ is the four-dimensional dilaton, the axion $\psi$ is the dual of the Kalb-Ramond 2-form in four dimensions,  $\mathcal{X}_{4}$ is the Gauss-Bonnet density \req{GB} and $\tensor{R}{_{\mu\nu\rho\sigma}}\tensor{\tilde{R}}{^{\mu\nu\rho\sigma}} $ is the Pontryagin density, with the dual Riemann tensor $\tensor{\tilde{R}}{^{\mu\nu\rho\sigma}}$ given by \req{dualriem}. The four-dimensional dilaton $\varphi$ is related to the ten-dimensional one by $\varphi=2(\hat\varphi-\hat\varphi_{\infty})$, where $\hat\varphi_{\infty}$ represents its asymptotic value, so that by construction $\varphi
\rightarrow 0$ at infinity. 

Now, under more general setups, one expects the dilaton to be stabilized with the inclusion of a potential $V(\varphi)$ that may be sourced, for example, from D-terms and nonperturbative corrections to the superpotential \cite{Cicoli:2013rwa}. The scalar potential may also be obtained from corrections to the Kahler potential and nonzero flux of $p$-forms. From a phenomenological point of view, the inclusion of all available effects then leads to more realistic physics \cite{Quevedo:2014xia}.  
On the other hand, the shift symmetry of the axion, which is associated to its origin as the dual of the Kalb-Ramond 2-form, prevents it from acquiring a mass.

It is indeed interesting to stabilize the dilaton, as the dilaton-Gauss-Bonnet sector of the theory \req{eq: heterotic action} is known to introduce singularities in extremal rotating black holes \cite{Kleihaus:2015aje,Chen:2018jed}. This is problematic for our purposes, since in that case the entropy and angular momentum become ill-defined, and therefore we must stabilize the dilaton in order to get a sensible result. 
Thus, in general we can assume that the theory \req{eq: heterotic action} can be supplemented with a potential $V(\varphi)$, which takes the form
\begin{align}
	V(\varphi) \sim V_{0} + \frac{1}{2}m^2 \varphi^2 + \dots .
\end{align}
when expanded around the vacuum expectation value of $\varphi$.  The effect of the dilaton will then be suppressed by the mass scale $m$, and as a first approximation we just can assume that the dilaton takes a constant value $\varphi=0$. However we need to account for the possible presence of a cosmological constant $2\Lambda=V_0$. In this scenario, the theory \req{eq: heterotic action} reduces to
\begin{align}\label{Istabilized}
	I_{\rm stabilized}=\frac{1}{16\pi G}\int d^4x\sqrt{|g|}\left[R-2\Lambda-\frac{1}{2}(\partial\psi)^2-\frac{\alpha'}{8}\psi\tensor{R}{_{\mu\nu\rho\sigma}}\tensor{\tilde{R}}{^{\mu\nu\rho\sigma}}+\frac{\alpha'}{8}\mathcal{X}_{4} \right]\, .
\end{align}
This is essentially dynamical Chern-Simons gravity \cite{CAMPBELL1991778,Alexander:2009tp}, except for the presence of the Gauss-Bonnet term. While this is a topological term that does not affect the equations of motion, it does make a contribution to the black hole entropy. 

From the point of view of effective field theory, this action is almost the most general we can have for a shift-symmetric pseudoscalar coupled to gravity. In fact, the only four-derivative operators that could be sourced are of the form $\psi\tensor{R}{_{\mu\nu\rho\sigma}}\tensor{\tilde{R}}{^{\mu\nu\rho\sigma}}$, $\mathcal{X}_4$, $R^2$, $R^{\mu\nu}R_{\mu\nu}$,  $R(\partial\psi)^2$ and $(\partial\psi)^4$. The last one is subleading since the solutions we study have $\psi=\mathcal{O}(\alpha')$, and therefore this term can be neglected. On the other hand, the three terms with Ricci curvature can be reabsorbed in a redefinition of the fields and of the couplings $G$, $\Lambda$ and $\alpha'$. Thus, only the two terms $\psi\tensor{R}{_{\mu\nu\rho\sigma}}\tensor{\tilde{R}}{^{\mu\nu\rho\sigma}}$ and $\mathcal{X}_4$ are relevant. The only possible generalization of \req{Istabilized} would be to consider an independent coupling (different from $\alpha'/8$) for the Gauss-Bonnet term. It is straightforward to generalize our results to that case, since the only effect of the Gauss-Bonnet density is to add a constant to the entropy. 

The equations of motion of this theory are given by
\begin{align} \label{eq: EE with corrections}
	\begin{split}
		R_{\mu\nu} - \frac{1}{2}g_{\mu\nu}R+\Lambda g_{\mu\nu} &= - \frac{\alpha'}{2} \nabla^\rho \nabla^\sigma\left(\tilde{R}_{\rho(\mu \nu) \sigma}\psi\right)+\frac{1}{2}\left(\partial_\mu \psi \partial_\nu \psi-\frac{1}{2} g_{\mu \nu}\left(\partial \psi\right)^2\right).
	\end{split}
	\\
	\begin{split}\label{eq: axion equation}
	\nabla^2\psi&=\frac{\alpha'}{8}\tensor{R}{_{\mu\nu\rho\sigma}}\tensor{\tilde{R}}{^{\mu\nu\rho\sigma}}
	\end{split}
\end{align}
We will study first the near-horizon geometries of \req{Istabilized} for $\Lambda=0$ and consider afterwards the effect of the cosmological constant, which significantly increases the complexity of the solutions. 

\subsection{Near-horizon geometry}
The first step in our construction of the solution is to analyze the near-horizon geometry, which enjoys an enhancement of symmetries, particularly $\mathrm{SO}(2,1)\times \mathrm{U}(1)$ symmetry. The metric ansatz is then given by
\begin{align} \label{eq: NH ansatz heterotic}
	ds^2 = g(x)ds^2_{\text{AdS}_{2}} + \frac{1}{f(x)}dx^2 + f(x) N(x)^2(d\phi - 2 a r dt)^2,
\end{align}
where
\begin{align} \label{eq: AdS2 metric}
	ds^2_{\text{AdS}_{2}} = -r^2 dt^2 + \frac{dr^2}{r^2},
\end{align}
is the metric of the AdS$_{2}$ space of unit radius. The parameter $a$ is associated to the angular momentum of the black hole and $x$ can be related to the polar coordinate $\theta\in [0,\pi]$ according to 
\begin{equation}
x=x_0\cos \theta
\end{equation}
for a certain $x_0$. In addition, the coordinate $\phi$ has a periodicity
\begin{equation}
\phi\sim \phi+\frac{2\pi}{\omega}\, ,
\end{equation}
for an undetermined constant $\omega$. 

The scalar field $\psi(x)$, the metric functions $f(x)$, $g(x)$ and $N(x)$ and the constants $x_0$ and $\omega$ have to be determined by solving the equations of motion.  These quantities can be expressed as a power series in $\alpha'$ and generally we can write
\begin{align}
\begin{split} \label{eq: psi f and g expansions}
	\Psi (x) = \sum_{n=0} \alpha^{\prime n} \Psi_n(x),
\end{split}
\end{align}
where $\Psi=\{\psi, f,g, N\}$ are the fields of the theory and
the subscript denotes the power of $\alpha^{\prime}$. Via the ansatz \eqref{eq: NH ansatz heterotic}, we can find the reduced Lagrangian of the theory 
 \begin{align}
L[\psi,f,g,N]=g(x)N(x)\mathcal{L}\Big|_{\text{ansatz}},
\end{align}
where $g(x)N(x)=\sqrt{-g}$, from which we can derive the equations of motion for $\Psi$
\begin{align}
\mathcal{E}_{\Psi}\equiv\frac{\delta L}{\delta \Psi}=\frac{\partial \mathcal{L}}{\partial \Psi} - \frac{\partial}{\partial x} \frac{\partial \mathcal{L}}{\partial \Psi'} + \frac{\partial^2}{\partial x^2} \frac{\partial \mathcal{L}}{\partial \Psi''} 
\, .
\end{align}
This is equivalent to solving the Einstein equations \req{eq: EE with corrections} and the scalar equation \req{eq: axion equation}. We note that our ansatz contains gauge freedom, since one of the three metric functions $f(x)$, $g(x)$ and $N(x)$ can be chosen at will (corresponding to a choice of $x$ coordinate).  For the solution at hand, we consider $N(x)$ to be a constant function that is exact in all orders of $\alpha'$, i.e.,
\begin{align} \label{eq: H function solution}
	N(x) = 1.
\end{align}
However, $N(x)$ must be included in the Lagrangian and its associated equation $\E_{N}$ must be computed before setting \eqref{eq: H function solution}. The reason for this is technical. We find that the Euler-Lagrange equations corresponding to $g(x)$ gives the $\mathcal{E}_{tt}$  and $\mathcal{E}_{rr}$ components of the Einstein equations (which are proportional to each other) while $N(x)$ yields $\mathcal{E}_{\phi\phi}$ and $f(x)$, a linear combination of $\mathcal{E}_{xx}$ and $\mathcal{E}_{\phi\phi}$. Therefore, by allowing $N(x)$ to enter into the ansatz, we extract $\mathcal{E}_{\phi\phi}$ and $\mathcal{E}_{xx}$, which imply the full set of Einstein equations due to the Bianchi identities. However, this is not true for other combinations.  For instance solving $\E_g$ and $\E_f$ does not imply the full Einstein's equations, because this is equivalent to solving $\E_{tt}$ and a linear combination of $\mathcal{E}_{xx}$ and $\mathcal{E}_{\phi\phi}$. This does not guarantee that the $\E_{xx}$ component is solved, since this equation always has a reduced order on account of the Bianchi identities. Thus, it is crucial to solve $\E_{xx}$ and this is the reason to include $N(x)$ in our ansatz, even though we impose \eqref{eq: H function solution}  eventually.

The reduced Lagrangian for the near-horizon extremal metric \eqref{eq: NH ansatz heterotic} is given by
\begin{align}
	\begin{split}
		L &= -2 f g N'' + N' \left(-3 g f'-2 f g'\right)
		\\&\quad
		-\frac{1}{2}g^{-1}N(-4 a^2 f N^2+2 g^2 f''+4 g f' g'+f g^2 \left(\psi '\right)^2+4 f g g''-f \left(g'\right)^2+4 g)
		\\&\quad 
		+\alpha'\left[\frac{1}{2}a g^{-2} N \psi  N' f \left(-16 a^2 f N^2+2 g^2 f''-2 f g g''+3 f \left(g'\right)^2+4 g\right)
		\right. \\&\quad\left. 
		+\frac{1}{2}a g^{-3} N^2 \psi  \left(g f'-f g'\right) \left(-8 a^2 f N^2+g^2 f''-g f' g'-f g g''+f \left(g'\right)^2+2 g\right)
		\right. \\&\quad\left. 
		+a f g^{-1} \psi  \left(N'\right)^2 \left(3 g f'-f g'\right)+N'' \left(2 a f^2 \psi  N'+a f g^{-1} N \psi  \left(g f'-f g'\right)\right)
		\right. \\&\quad \left. 
		+\frac{1}{2}a g^{-2} N \psi  N'\left(3 g^2 \left(f'\right)^2-6 f g f' g'\right)\right]\, ,
	\end{split}
\end{align}
plus a total derivative term, coming from the Gauss-Bonnet density, which we are discarding. 
When we insert the expansion \req{eq: psi f and g expansions} the equations of motion can be studied order by order in $\alpha'$ and contain subleading terms
\begin{align}
	\mathcal{E}_{\Psi} = \mathcal{E}_{\Psi}^{(0)}+ \alpha' \mathcal{E}_{\Psi}^{(1)} + \alpha^{\prime 2} \mathcal{E}_{\Psi}^{(2)}\,  + \dots \,.
\end{align}
This leads to the following equations of motion $\mathcal{E}_{\Psi}^{(0)}$ at zeroth order
\begin{align}
	\begin{split}
		\E_{\psi}^{(0)} &= g_0 f_0'  \psi_0 ' +f_0  g_0'  \psi_0 ' +f_0  g_0  \psi_0 ''=0 \, ,
		\\
		\E_{g}^{(0)}  &= \frac{2 a^2}{g_0 }-g_0'' +\frac{(g_0')^2}{2 g_0 }-\frac{1}{2} g_0  (\psi_0')^2=0\, ,
		\\
		\E_{f}^{(0)} &=-\frac{2 a^2 f_0 }{g_0 ^2}-f_0'' -\frac{f_0'  g_0' }{g_0 }-\frac{f_0  g_0'' }{g_0 }+\frac{f_0  (g_0')^{2}}{2 g_0 ^2}-\frac{1}{2} f_0  (\psi_0')^2=0\, ,
		\\
		\E_{N}^{(0)} &= \frac{6 a^2 f_0}{g_0}-f_0' g_0'-2 f_0 g_0''+\frac{f_0 \left(g_0'\right)^2}{2 g_0}-\frac{1}{2} f_0 g_0 \left(\psi_0 '\right)^2-2=0\, ,
	\end{split}
\end{align}
where we have already imposed \eqref{eq: H function solution}. Note that, indeed, the equation $\E_{N}^{(0)}$ is only of first order in $f_0$, and $\E_{f}$ is actually implied by the rest of the equations. 

By setting $\psi_0=0$ and demanding that both $f_0(x)$ and $g_0(x)$ be even functions, we find the NHEK solution, which takes the form
\begin{align} \label{eq: psi0 f0 g0}
	\psi_0(x) &= 0, \quad f_0(x) = \frac{a^2-x^2}{a^2+x^2}, \quad g_0(x) = x^2 + a^2\, . 
\end{align}
Note that this satisfies 
\begin{align}
	f_0(a)=f_0(-a)=0\, ,
\end{align}
so that $x_0=a$ in this case. In addition, absence of a conical singularity at the poles $x=\pm x_0$ requires the following relation for the period of $\phi$ 
\begin{equation}
f'(x_0)=-f'(-x_0)=-2\omega\, ,
\end{equation}
from where we get $\omega=1/(2a)$.

We now look at the $\alpha'$ corrections. The scalar $\psi$ receives corrections at first order and the equation $\E_{\psi}^{(1)}$ reads
\begin{align} \label{eq: psi1}
\begin{split}
	0 &= \left(a^2-x^2\right)\psi_{1}''-2 x\psi_{1}'+\frac{12 a^3 \left(a^2-3 x^2\right) \left(3 a^2 x-x^3\right)}{\left(a^2+x^2\right)^5}\, .
\end{split}
\end{align}

For a solution of $\psi$ that is corrected at order $\alpha'$, the right hand side of the Einstein equations \req{eq: EE with corrections} is $\mathcal{O}(\alpha'^{2})$. This means that at order $\alpha'$, only the vacuum solution is to be satisfied for $f$ and $g$ and therefore can we can set  $f_{1}=g_{1}=0$ (a different choice would simply correspond to a redefinition of the integration constants of the zeroth-order solution). Therefore, the first non-zero higher derivative corrections for $f(x)$ and $g(x)$ are at $\mathcal{O}(\alpha'^{2})$. At this order, the equations of motion $\mathcal{E}^{(2)}_{g}$ and $\mathcal{E}^{(2)}_{N}$ are\footnote{As before, we do not need $\mathcal{E}^{(2)}_{f}$ since it is redundant.}
\begin{align} \label{eq: g2}
	\begin{split}
	0 &=\frac{1}{2 \left(a^2+x^2\right)^2}\left(\left(a^2+x^2\right) \left(-\left(a^2+x^2\right) \left(\left(a^2+x^2\right)(\psi_{1}')^2+2 g_{2}'' \right)+4 x g_{2}' -4 g_{2} \right)\right. \\&\left. \quad\quad\quad\quad\quad\quad\quad +2 a x \left(x^2-3 a^2\right)\psi_{1}'' \right)\, ,
\end{split}
\end{align}
\begin{align}
\begin{split}\label{eq: f2}
	0&= \frac{2 f_2 \left(a^2-x^2\right)}{a^2+x^2}-2 x f_2'-\frac{2 g_2 \left(3 a^4-2 a^2 x^2-x^4\right)}{\left(a^2+x^2\right)^3}+\frac{2 \left(3 a^2 x-x^3\right) g_2'}{\left(a^2+x^2\right)^2}-\frac{2 \left(a^2-x^2\right) g_2''}{a^2+x^2}\\&-\frac{1}{2} \left(a^2-x^2\right) \left(\psi _1'\right){}^2
	\quad +\frac{2 a \left(x^3-3 a^2 x\right)^2 \psi _1'}{\left(a^2+x^2\right)^4}-\frac{2 a x \left(3 a^4-4 a^2 x^2+x^4\right) \psi _1''}{\left(a^2+x^2\right)^3}\, ,
\end{split}
\end{align}
where we have imposed the zeroth order solution \eqref{eq: psi0 f0 g0} and \eqref{eq: H function solution}. For purposes of clarity, from now on, we define the $x$ coordinate such that 
\begin{align}
	x = a y\, ,
\end{align}
and suppress the arguments of the functions, which are now functions of $y$. Similarly, $x_0=a y_0$. Then, the solution to \eqref{eq: psi1}, \eqref{eq: g2}, and \eqref{eq: f2} is found to be\footnote{This solution (expressed in different coordinates) was already obtained in \cite{Chen:2018jed}. The same reference also computed the black hole entropy and expressed it in terms of a parameter claimed to be the mass. However,  such identification is not justified since one cannot obtain the mass of the black hole from the NHEG, as we discuss below.}
\begin{align}
	\begin{split} \label{eq: psi1 solution heterotic}
		\psi_{1} &= -\frac{\arctan y}{2 a^2} -\frac{y \left(y^4+2 y^2-7\right)}{4 a^2 \left(y^2+1\right)^3}\, ,
	\end{split} 
	\\
	\begin{split} \label{eq: f2 solution heterotic}
		f_{2} &= -\frac{ y \left(3 y^2+43\right) \arctan y}{32 a^4 \left(y^2+1\right)^2}\\&\quad -\frac{\left(315 y^{12}+5495 y^{10}+21966 y^8+31030 y^6+51263 y^4+963 y^2+1192\right)}{3360 a^4 \left(y^2+1\right)^7}\, ,
	\end{split}
	\\
	\begin{split} \label{eq: g2 solution heterotic}
		g_{2} &= \frac{5 y \arctan y}{8 a^2} + \frac{4095 y^{10}+19285 y^8+35014 y^6+35610 y^4-5861 y^2+241}{6720 a^2 \left(y^2+1\right)^5}\, ,
	\end{split}
\end{align}
where we demanded $\psi_1$ to be an odd function of $y$ --- since it is a pseudoscalar --- and $f_2$ and $g_2$ to be even functions. In this process there is one integration constant left, but this can be seen to be equivalent to a redefinition of $a$, and hence we have set it to zero. 
The full solution is then parametrized by the parameter $a$ that is associated with the angular momentum. The solution is regular at the poles --- it does not have divergencies --- which are now shifted to $x=\pm x_0=\pm a y_0$, with 
\begin{align}
y_{0} = 1 + \frac{\alpha'^2}{a^4}\delta y_{}\, .
\end{align}
We can solve for the shift $\delta y$ from 
\begin{align} \label{eq: delta y solution heterotic}
	f(x_{0}) = 0 \qquad \to \qquad \delta y  = \frac{-334-115 \pi }{1280} \, ,
\end{align}
which also allows us to solve for $\omega$ from the condition that no conical defects arise at the poles
\begin{align} \label{eq: omega solution heterotic}
	f'(x_{0}) = -2\omega, \quad \to \quad \omega &= \frac{1}{2 a}+\alpha'^2 \left(\frac{1972+315 \pi }{53760 a^5}\right).
\end{align}

\subsection{Entropy and angular momentum}\label{subsec:SJheterotic}
\subsubsection*{Entropy}
With the solution of the near-horizon geometry, we can compute the conserved charges of the theory. We consider the variation of the Lagrangian with respect to the Riemann tensor as defined in \eqref{eq: def of P} for the Lagrangian \eqref{eq: heterotic action} and we find 
\begin{align} \label{eq: P for heterotic theory}
\bar{P}^{\mu \nu \rho \sigma}= \frac{1}{16\pi G}\left[ g^{\mu[\rho} g^{\sigma] \nu}-\frac{\alpha^{\prime}}{8} \psi\left(\tilde{R}^{\mu \nu \rho \sigma}+\tilde{R}^{\rho \sigma \mu \nu}-\tilde{R}^{\mu [\nu \rho \sigma]}-\tilde{R}^{[\rho\sigma| \mu |\nu]}\right)-\frac{\alpha'}{4}\tilde{\tilde{R}}^{\mu\nu\rho\sigma}\right]\, ,
\end{align}
where
\begin{equation}
\begin{aligned}
\tensor{\tilde{\tilde{R}}}{^{\mu\nu}_{\rho\sigma}}=\frac{1}{4}\epsilon^{\mu\nu\alpha\beta}\epsilon_{\rho\sigma\lambda\tau}\tensor{R}{_{\alpha\beta}^{\lambda\tau}}=-\tensor{R}{^{\mu\nu}_{\rho\sigma}}+4\tensor{R}{^{[\mu}_{ [\rho}}\tensor{\delta}{^{\nu]}_{ \sigma]}}-R\tensor{\delta}{^{[\mu}_{ [\rho}}\tensor{\delta}{^{\nu]}_{ \sigma]}} 
\end{aligned}
\end{equation}
comes from the variation of the Gauss-Bonnet density.
We note that this indeed satisfies the algebraic Bianchi identity $\bar{P}^{\mu [\nu \rho \sigma]}=0$.
For the computation of Iyer-Wald's entropy \eqref{eq: IW entropy formula} we need the binormal to the horizon, which for the metric \eqref{eq: NH ansatz heterotic} is given by
\begin{equation}
\epsilon_{tr}=-\epsilon_{rt}=g\, .
\end{equation} 
We then have (setting $N=1$)
\begin{align}
S &= -8\pi a \int dy d\phi g^2\bar{P}^{trtr}\, ,
\end{align}
and we only need the $\bar{P}^{trtr}$ component, which reads
\begin{align} \label{eq: Ptrtr for heterotic theory}
\bar{P}^{trtr}= & \frac{1}{32\pi G}\left[ -\frac{1}{g^2}+ \alpha' \frac{\psi}{2a^2g^3}\frac{d}{dy}\left(\frac{f}{g}\right) +\frac{\alpha'}{4a^2 g^2}\frac{d^2f}{dy^2}\right]\, .
\end{align}
Imposing the near-horizon solution via \eqref{eq: psi1 solution heterotic}, \eqref{eq: f2 solution heterotic}, and \eqref{eq: g2 solution heterotic}, the black hole entropy is\footnote{Note that the contribution from the Gauss-Bonnet is a total derivative and in order to evaluate it we only need to use that $\tfrac{df}{dy}(\pm y_0)=\mp 2 a\omega$, imposed from the boundary conditions at the poles. }
\begin{align}
\begin{split}
	S &= \frac{a}{16G} \int dy d\phi \left(4-\frac{\alpha'}{a^2}\frac{d^2f}{dy^2}+\frac{2 \alpha'^2 y \left(y^2-3\right) \arctan y}{a^4 \left(y^2+1\right)^3}+\frac{\alpha'^2 y^2 \left(y^2-3\right) \left(y^4+2 y^2-7\right)}{a^4 \left(y^2+1\right)^6}\right),
\end{split}
\end{align}
where we evaluate the integral from $- y_{0} \leq y \leq y_{0}$ and $0 \leq \phi \leq \frac{2\pi}{\omega}$, which gives us

\begin{align}\label{entropy heterotic functio of a}
\begin{split}
	S &= \frac{\pi  a}{G \omega }+\alpha'\frac{\pi}{2G} + \alpha'^2 \frac{\pi \left(3840 \delta y +45 \pi +64\right)}{3840 a^3 G \omega } + \mathcal{O}(\alpha'^3) \\
	&= \frac{2 \pi  a^2}{G}+\alpha'\frac{\pi}{2G}-\alpha'^2 \frac{\pi  \left(805 \pi +2846\right)}{4480 a^2 G}+O\left(\alpha'^3\right),
\end{split}
\end{align}
The values of $\omega$ from \eqref{eq: omega solution heterotic} and $\delta y$ from \eqref{eq: delta y solution heterotic} have been imposed in the second equal sign.

\subsubsection*{Angular Momentum}
In order to obtain the physically meaningful relation $S(J)$, we must also identify the angular momentum of the solution. We start with a general Killing vector

\begin{equation}
\xi=\xi^{0}\partial_t+\xi^{3}\partial_{\phi}\, ,
\end{equation}
which as a 1-form reads
\begin{align} \label{eq: general Killing vector}
	\boldsymbol{\xi} 
	&=
	f (2 a \xi^{0} r-\xi^{3}) (2 a r dt-d\phi)-\xi^{0} r^2 g dt\, .
\end{align}
The Killing vector that generates $2\pi$ rotations corresponds to the choice $\xi^{0}=0$, $\xi^{3}=\omega^{-1}$, but for now we will leave these constants to be arbitrary for the sake of generality.

In order to utilize the Komar integral \eqref{eq: def of J}, we are required to compute a 2-form $\boldsymbol{\Omega}$ satisfying \req{omegaformequation}. The Hodge dual of \eqref{eq: general Killing vector} reads
\begin{align}
	\star \boldsymbol{\xi} &= 
	a g (\xi^{3} dr\wedge dt\wedge dy+\xi^{0} dr\wedge dy\wedge d\phi)\, ,
\end{align}
and since the Lagrangian depends only on the coordinate $y$, one possible solution to $\boldsymbol{\Omega}$ takes the form
\begin{align}\label{eq:omegaforstringy}
	\boldsymbol{\Omega} &= -\frac{1}{2} a g r (\xi^{3} dt\wedge dy+\xi^{0} dy\wedge d\phi)\mathcal{L}.
\end{align}
At this point, let us address the issue of the ambiguity in $\boldsymbol{\Omega}$. To this result we can add an exact form --- particularly a term of the form $dy\wedge d\phi$ --- which could contribute to the total charge. As we discussed in section~\ref{sec:charges}, this ambiguity is fixed by the condition that the integral of $\boldsymbol{\Omega}$ at infinity vanishes. Even though we do not know the black hole solution beyond the near-horizon region, the global geometry of a stationary and axisymmetric black hole can always be written in the form
\begin{equation}
ds^2=g_{tt}dt^2+2g_{t\phi}dt d\phi+g_{rr}dr^2+g_{yy}dy^2+g_{\phi\phi} d\phi^2\, ,
\end{equation} 
where all the components are functions of $r$ and $y$ only. Then, for the rotational Killing vector, $\xi=\xi^{3}\partial_{\phi}$, $\boldsymbol{\Omega}$ must satisfy
\begin{align}
\diff\boldsymbol{\Omega}=-\frac{1}{2}\star \boldsymbol{\xi} \, \mathcal{L}= 
-\frac{1}{2}\sqrt{|g|} \mathcal{L} \xi^{3} dr\wedge dt\wedge dy\, ,
\end{align}
and hence, since $\sqrt{|g|} \mathcal{L}$ is independent of $\phi$, we can choose this $2$-form as
\begin{equation}
\boldsymbol{\Omega}=-\frac{1}{2} \xi^{3} dt\wedge dy \int dr \sqrt{|g|} \mathcal{L}\, .
\end{equation}
The integral of this $\boldsymbol{\Omega}$ on any constant $t$ and $r$ surface --- including the sphere at infinity --- vanishes because $\boldsymbol{\Omega}$ does not have a $dx\wedge d\phi$ component. On the other hand, it is clear that in the near horizon region this choice of $\boldsymbol{\Omega}$ corresponds precisely to the $\xi^{3}$ component of \req{eq:omegaforstringy}. Therefore, \req{eq:omegaforstringy} is the correct choice in order to compute the angular momentum.

Since the Komar integral \req{eq: def of J} is independent of the chosen $\Sigma$, we choose a surface of constant $t$ and $r$ and we find the relevant term in the Noether-Komar charge 2-form \req{Komar2form} to be given by
\begin{align}
	\begin{split} \label{eq: integrand J heterotic}
		&-\boldsymbol{\epsilon}_{\mu\nu}\bar{P}^{\mu\nu\alpha\beta}\nabla_{\alpha}\xi_{\beta}-2\nabla_{\beta}\bar{P}^{\mu\nu\alpha\beta}\xi_{\alpha}-\boldsymbol{\Omega} = \frac{1}{16\pi G} \bigg[ a dy \wedge d\phi \left(r\xi^{0} W_0 + \xi^{3}W_{3}\right) + \dots\, \bigg],
	\end{split}
\end{align}
where
\begin{align}
\begin{split}
W_{0}&=  \frac{a^2f}{g}+\frac{1}{2a^2} g f''+ \frac{1}{a^2} f' g'+ \frac{1}{4a^2} f g \left(\psi '\right)^2+\frac{1}{a^2}  f g''- \frac{1}{4a^2g } f \left(g'\right)^2\\
&\quad+\frac{\alpha'}{24 a^3 g^3}  \left(\psi ' \left(8 a^3 f g \left(4 a^2 f-g\right)-4 a f g^3 f''+4 a f^2 g^2 g''-4 a f^2 g \left(g'\right)^2 
\dvvv \quad
+4 a f g^2 f' g'\right)+\psi  \left(g' \left(4 a^3 f \left(8 a^2 f-g\right)+4 a g^2 \left(f'\right)^2\right)-4 a^3 g f' \left(8 a^2 f-g\right)
\dvvv \quad +4 a f g^2 \left(f g^{(3)}-g f^{(3)}\right)+8 a f^2 \left(g'\right)^3-12 a f g f' \left(g'\right)^2+g'' \left(12 a f g^2 f'-12 a f^2 g g'\right)
\dvvv \quad +f'' \left(4 a f g^2 g'-4 a g^3 f'\right)\right)\right) \, ,
\end{split}
\\ \nonumber \\
\begin{split}
W_{3} &= - \frac{af}{g} + \frac{\alpha'}{24 a^3 g^3}  \left(\psi ' \left(-4 a^2 f g \left(4 a^2 f-g\right)+2 f g^3 f''-2 f^2 g^2 g''+2 f^2 g \left(g'\right)^2 -2 f g^2 f' g'\right)
\dvv \quad +\psi  \left(g' \left(2 a^2 f \left(g-20 a^2 f\right)+g^2 \left(f'\right)^2\right)+2 a^2 g f' \left(20 a^2 f-g\right)+2 f g^2 \left(g f^{(3)}-f g^{(3)}\right)
\dvvv \quad -f^2 \left(g'\right)^3+g'' \left(3 f^2 g g'-3 f g^2 f'\right)+f'' \left(f g^2 g'-g^3 f'\right)\right)\right) \, ,
\end{split}
\end{align}
where the primes indicate derivatives with respect to $y$ and the ellipses denote the other components of the 2-form.
We must then impose the solution of $\psi$, $f$ and $g$ with their corresponding $\alpha'$ corrections found in \eqref{eq: psi1 solution heterotic}, \eqref{eq: f2 solution heterotic} and \eqref{eq: g2 solution heterotic}. 

The integrand \eqref{eq: integrand J heterotic} must be integrated from $- y_{0} \leq y \leq y_{0}$ and $0 \leq \phi \leq \frac{2\pi}{\omega}$. We observe that the term proportional to $\xi^{0}$, coming from the time-like Killing vector is proportional to $r$. However, by construction the Komar integral should be independent of the integration surface. In fact, one can check that the integration of the $\xi^{0}$ term in the interval $- y_{0} \leq y \leq y_{0}$ vanishes at order $\alpha'^2$. The charge associated to $\xi^{0}$ would be the mass of the black hole, but as this result shows, it is not possible to obtain it from the near-horizon geometry. Ultimately, the reason for this follows from the fact that one cannot identify the asymptotic time coordinate from the near-horizon geometry due to its invariance under rescalings $(t,r)\rightarrow (\gamma t , r/\gamma)$.

The angular momentum is obtained from the integral proportional to $\xi^{3}$ with the choice $\xi^{3}=1/\omega$. Once the dust settles, we get the result
\begin{align}
\begin{split}
	J &=-2\int_{\Sigma}\left[\boldsymbol{\epsilon}_{\mu\nu}\left(\bar{P}^{\mu\nu\alpha\beta}\nabla_{\alpha}\xi_{\beta}+2\nabla_{\beta}\bar{P}^{\mu\nu\alpha\beta}\xi_{\alpha}\right)+\boldsymbol{\Omega}\right] = \frac{1}{4\left(G \omega ^2\right)}-\frac{\alpha'^2 \left(245 \pi +437\right)}{17920 a^4 G \omega ^2}+O\left(\alpha'^3\right) \, .
\end{split}
\end{align}
Using  the value of $\omega$ \req{eq: omega solution heterotic}, this can be expressed in terms of $a$ alone
\begin{align}
	\begin{split} \label{eq: J with corrections heterotic}
		J &= 
		\frac{a^2}{G} -\alpha'^2\frac{469+150 \pi}{1920 a^2G }+O\left(\alpha'^3\right)\, ,
	\end{split}
\end{align}
which in turn allows us to express $a$ as a function of $J$,
\begin{align}
\begin{split} \label{eq: a with corrections heterotic}
a^2 &= G J +\alpha'^2\frac{469+150 \pi}{1920 G J}+O\left(\alpha'^3\right)\, ,
\end{split}
\end{align}

\subsubsection*{The relation $S(J)$}
Combining the result for the entropy \req{entropy heterotic functio of a} together with \req{eq: a with corrections heterotic}, we obtain the entropy as a function of the angular momentum
\begin{align} \label{eq: entropy for heterotic string}
		S = 2 \pi  J+\alpha'\frac{\pi}{2G}-\alpha'^2 \frac{\pi}{JG^2}\left(\frac{493 }{3360}+\frac{3 \pi}{128}\right)+O(\alpha'^{3})\, .
\end{align}
Observe that, while the first-order correction coming from the topological Gauss-Bonnet contribution is positive, the $\mathcal{O}(\alpha'^2)$ correction is negative. Thus, the higher derivative corrections from the Pontryagin density in \req{Istabilized} lead to a smaller black hole entropy. 

\subsection{Adding a cosmological constant}

Motivated by the stabilization of the dilaton as mentioned in section~\ref{subsection: the four dimensional heterotic action}, we consider the theory \req{Istabilized} with $\Lambda\neq 0$ and study how the near-horizon geometry and the conserved charges are affected. 

The solution to the near-horizon geometry can still be found analytically. At leading order in $\alpha'$, the functions in the metric $g_0$ and $N$ take their form in \eqref{eq: psi0 f0 g0} and \eqref{eq: H function solution} and the scalar is still zero. However, $f_0$ is now given by
\begin{align}
	f_{0} &= \frac{1-y^2}{1+y^2}-\frac{a^2 \Lambda  \left(y^4+6 y^2-3\right)}{3 \left(1+y^2\right)} \, ,
\end{align}
where the second term on the right hand side is proportional to the cosmological constant. Moreover, $y_0$ --- determined by $f_0(y_0)=0$ --- is no longer equal to unity and can be solved via the quartic relation
\begin{align}
	\Lambda &=  -\frac{3 \left(y_0^2-1\right)}{a^2 \left(y_0^4+6 y_0^2-3\right)} \, .
\end{align}
It is indeed useful to express the solution in terms of $y_0$ instead of $\Lambda$, as we do next. The first nonzero $\alpha'$ corrections for $\psi, f,$ and $g$ all have a similar functional dependence on $y$ and they are given by
\begin{align}
\begin{split} \label{eq: psi1 with lambda heterotic}
	\psi_{1} &= \frac{1}{4 a^2 \left(y_{0}^4+6 y_{0}^2-3\right)}\left\{-2\arctan y \left(y_{0}^4+4 y_{0}^2-1\right)- \eta \left(y_{0}^2-3\right)^2\operatorname{arctanh} \left(y \eta \right)
	\right. \\& \left. \quad\quad\quad\quad\quad\quad\quad\quad\quad\quad\quad\quad\quad +  \frac{y \left(7 y_{0}^4+2 y_{0}^2+19\right)-y^3 \left(y^2+2\right) \left(y_{0}^4+14 y_{0}^2-11\right)}{\left(y^2+1\right)^3} \right\},
\end{split}
\\ \nonumber \\
\begin{split} \label{eq: g2 with lambda heterotic}
	g_{2} &= \frac{1}{32 a^2 \left(y_{0}^4+6 y_{0}^2-3\right)^2}\left\{2y \left(10 y_{0}^8+29 y_{0}^6+25 y_{0}^4+99 y_{0}^2-3\right) \arctan y
	\right. \\&\quad \left.
	+ 3 y \frac{\left(y_{0}^2-3\right)^2}{(y_0^2+3)} \eta \left(y_{0}^6+3 y_{0}^4-5 y_{0}^2+9\right) \operatorname{arctanh}\left(y \eta\right)
	\right. \\&\quad\left.
	+\left(y^2-1\right) y_{0}^2 \left(y_{0}^2-3\right)^2 \left(y_{0}^2-1\right) \log \left[\tfrac{y_{0}^2+3-y^2 \left(y_{0}^2-1\right)}{y^2+1}\right]
	+\frac{\sum_{n=0}^{5} \gamma_{g,2n}y^{2n}}{105 \left(y^2+1\right)^5 \left(y_{0}^2+3\right)}
\right\} \, ,
\end{split}
\\ \nonumber \\
\begin{split} \label{eq: f2 with lambda heterotic}
	f_{2} &= \frac{1}{32 a^4 (1 + y^2)^2 (-3 + 6 y_0^2 + y_0^4)^3}\\&\times\left\{y\arctan y\sum_{n=0}^{2}\tau_{f,(2n)}y^{2n}
	+y\frac{\left(y_{0}^2-3)\right)^2}{(y_0^2+3)}\eta \arctan \left(y \eta \right) \sum_{n=0}^{2} \kappa_{f,2n}y^{2n}
	\right. \\& \left. \quad + y_{0}^2 \left(y_{0}^2-3\right)^2 \left(y_{0}^2-1\right)\log \left[\tfrac{y_{0}^2+3-y^2 \left(y_{0}^2-1\right)}{y^2+1}\right]\sum_{n=0}^{2} \rho_{f,2n}y^{2n}
	 -\frac{\sum_{n=0}^{7}\gamma_{f,2n}y^{2n}}{150(1+y^2)^{5}(y_0^2+3)}\right\},
\end{split}
\end{align}

where we have defined
\begin{align}
	\eta &= \sqrt{\frac{1-y_{0}^2}{y_{0}^2+3}}.
\end{align}
The various constants $\tau, \kappa, \rho$ and $\gamma$ can be found in appendix~\ref{appendix: heterotic string}. Moreover, the shift in $y_0$ (such that $f(y_0+\alpha'^2 \delta y_0/a^4)=0$) due to the higher derivative terms is given by 
\begin{align}
	\begin{split}
		\delta y_{0} &= \frac{1}{(1 + y_{0}^2) (-3 + 6 y_{0}^2 + y_{0}^4)^2} \\&\quad \times
		\left\{\frac{1}{64(y_0^2-3)}\arctan y_0\sum_{n=0}^{6}\tau_{\delta,2n}y^{2n}_0
		+ \frac{1}{32}\frac{y_{0}^2-3}{y_0^2+3}\eta \arctan \left(y_0 \eta \right) \sum_{n=0}^{5} \kappa_{\delta,2n}y_0^{2n}
		\right. \\& \left. \quad - \frac{1}{64}y_{0}^2 \left(y_{0}^2-3\right) \left(y_{0}^2-1\right)\log \left[\tfrac{y_{0}^2+3-y^2_0 \left(y_{0}^2-1\right)}{y^2_0+1}\right]\sum_{n=0}^{3} \rho_{\delta,2n}y^{2n}_0
		-\frac{\sum_{n=0}^{7}\gamma_{\delta,2n}y^{2n}_0}{6720 y_0(y^2_0-3)(y_0^2+3)}\right\},
	\end{split}
\end{align}
where
\begin{align}
	\begin{split}
		\tau_{\delta,2n} &= \{-281,348,-505,-480,-475,-92,13\}, 
		\\
		\kappa_{\delta,2n} &= \{-81,81,42,-66,-9,1\},
		\\
		\rho_{\delta,2n} &= \{3,-3,9,-1\},
		\\
		\gamma_{\delta,2n} &= \{-9180,-71019,-19890,-55693,-234600,-72989,-5250,1365\}.
	\end{split}
\end{align}
With the solution at hand, we can now investigate the effect of the cosmological constant on the conserved charges. We start with the same general Killing vector \eqref{eq: general Killing vector} and we see that $\boldsymbol{\Omega}$ takes the similar form as \eqref{eq:omegaforstringy} except that the Lagrangian now includes a nonzero cosmological constant.  Following the procedure of evaluating the Komar integral \eqref{eq: def of J} and the Iyer-Wald entropy \eqref{eq: IW entropy formula}, the conserved charges of the theory are
\begin{align} \label{eq: J with Lambda leading term heterotic}
	J & = \frac{3 y_0 \left(1-y_0^2\right)}{G \Lambda  \left(3-y_0^2\right)^2}  + \mathcal{O}(\alpha'^2) \, ,
	\\
	S &=\frac{3 \pi  \left(1-y_0^2\right)}{G \Lambda  \left(3-y_0^2\right)}  + \mathcal{O}(\alpha')\, ,
\end{align}
where the $\alpha'$ corrections take a quite involved form that we are not showing yet. 
Note that even in Einstein gravity one cannot easily obtain the relation $S(J)$ explicitly and it is more convenient to express the final results in terms of the parameter $y_0$. Now, both $S$ and $J$ receive $\alpha'$ corrections which depend on $y_0$, but we can simplify these relations by using a different parameter $\zeta$ defined as
\begin{align}
y_0 = \zeta + \alpha^{\prime 2} \delta \zeta +\mathcal{O}(\alpha'^3)\, .
\end{align}
By choosing $\delta \zeta$ appropriately, the angular momentum can be recast in the same form as in Einstein gravity \eqref{eq: J with Lambda leading term heterotic}, \textit{i.e.},
\begin{align} \label{eq: J with Lambda with corrections heterotic}
	J &= \frac{3 \zeta  \left(1-\zeta ^2\right)}{G \Lambda  \left(3-\zeta ^2\right)^2}\, .
\end{align}

Using this parameter $\zeta$, the entropy reads
\begin{align}\label{eq: S with Lambda zeta heterotic}
	\begin{split}
		S &= \frac{3 \pi  \left(1-\zeta ^2\right)}{G\Lambda  \left(3-\zeta ^2\right)}+\alpha'\frac{\pi}{2G}
		+\alpha^{\prime 2} \frac{\pi \Lambda}{G}\left\{-\frac{\left(-3+\zeta ^2\right)^4 \left(1+\zeta ^2\right)  \arctan\left(\zeta  \sqrt{\frac{1-\zeta ^2}{3+\zeta^2}}\right)}{192 \zeta  \left(-3+6 \zeta ^2+\zeta ^4\right) \sqrt{3-\zeta ^2 \left(2+\zeta ^2\right)}}
		\right. \\& \left. \quad
		-\frac{8925+71155 \zeta ^2-104972 \zeta ^4+75970 \zeta ^6+7161 \zeta ^8+4235 \zeta ^{10}+630 \zeta ^{12} }{20160 \left(-3+\zeta ^2\right) \left(-1+\zeta ^2\right) \left(1+\zeta ^2\right) \left(-3+6 \zeta ^2+\zeta ^4\right)}
		\right. \\& \left. \quad 
		-\frac{\left(-166+675 \zeta ^2-1009 \zeta ^4+550 \zeta ^6+112 \zeta ^8+23 \zeta ^{10}+7 \zeta ^{12}\right) \arctan \zeta }{192 \zeta  \left(-3+\zeta ^2\right) \left(-1+\zeta ^2\right) \left(-3+6 \zeta ^2+\zeta ^4\right)}
		\right\} +O\left(\alpha^{\prime 3}\right) .
	\end{split}
\end{align}
The two expressions \req{eq: J with Lambda with corrections heterotic} and \req{eq: S with Lambda zeta heterotic} allow us to study the relation $S(J)$ parametrically. 
We conclude this section by studying two limits in which the entropy can be expressed explicitly in terms of the angular momentum. For small black holes --- that is, those with $J \ll 1/(G|\Lambda|)$ --- we find that $\zeta$ can be expressed in terms of a series expansion in $J$
\begin{align}
	\zeta &= 1-\frac{2}{3} (J G) \Lambda -\frac{14}{9} (J G)^2 \Lambda ^2 + \mathcal{O}(J^3),
\end{align}
and therefore, we can express the entropy as 
\begin{align}
	\begin{split}
	S(J) &= 2\pi J+\frac{8\pi}{3} G\Lambda J^2+\alpha'\frac{\pi}{2G} 
 +\alpha^{\prime 2} \frac{\pi}{G}\bigg[-\frac{ (1972+315 \pi )}{13440 (GJ)}+\frac{(4912+1785 \pi ) \Lambda }{13440}	\\&\quad-\frac{(18586+2625 \pi ) (JG)\Lambda ^2}{30240}-\frac{(118432+2135 \pi ) (JG)^2\Lambda ^3}{60480}\bigg]+\mathcal{O}(\alpha'^3, J^3)
	\end{split}
\end{align}
Note that the leading terms in the $\alpha'$ expansion reproduces what we have computed previously in \eqref{eq: entropy for heterotic string} for asymptotically flat black holes. We also note that there is a term constant in $J$ at order $\mathcal{O}(\alpha^{\prime 2})$ that would yield a constant shift to the overall entropy of small black holes.

For large black holes, $J \gg 1/(G|\Lambda|)$, the expansion takes the form
\begin{align}
	\zeta &= \sqrt{3}+\frac{\sqrt[4]{3}}{\sqrt{-2JG\Lambda}}-\frac{3}{4\Lambda JG}+\mathcal{O}\left(\frac{1}{J}\right)^{3/2},
\end{align}
such that the entropy is given by
\begin{align}\label{SstringylargeJ}
	S(J) &= \pi\left(\frac{\sqrt{12}J}{G|\Lambda|}\right)^{1/2}\left[1  - \alpha^{\prime 2}\Lambda^2\left(\frac{983}{6720}+\frac{107\pi}{648\sqrt{3}}\right)\right]+\mathcal{O}\left(1\right) \, .
\end{align}
This limit only exists in the case of a negative cosmological constant, as black holes in de Sitter space cannot be arbitrarily large on account of the cosmological horizon. 
\section{Cubic gravity}
\label{sec:cubic}
We now push forth in our analysis of higher derivative corrections by considering a different theory, namely the following six-derivative extension of GR
\begin{equation}\label{cubictheory}
I_{\rm cubic}=\frac{1}{16\pi G}\int d^4x\sqrt{|g|}\left\{-2\Lambda+R+\lambda \mathcal{P}+\tilde\lambda \tilde{\mathcal{P}}\right\}\, ,
\end{equation}
where $\mathcal{P}$ and $\tilde{\mathcal{P}}$ are the even-parity and odd-parity cubic densities given by \req{ECG} and \req{EvilECG}. Also, $\lambda$ and $\tilde\lambda$ are couplings with dimensions of length to the fourth.   
Clearly, this Lagrangian provides a general effective field theory extension of GR to six derivatives. In fact, one can always get rid of higher-derivative terms with Ricci curvature via field redefinitions \cite{Cano:2019ore,Bueno:2019ltp}, so the theory can be recast into the more standard form\footnote{To arrive to this result one also must take into account that the two different Riemann cube contractions in \req{ECG} can be related through the identity $\tensor{R}{_{\mu\nu}^{[\rho\sigma}}\tensor{R}{_{\rho\sigma}^{\alpha\beta}}\tensor{R}{_{\alpha\beta}^{\mu\nu]}}=0$. }

\begin{align}
\begin{split}
I_{\rm cubic}=\frac{1}{16\pi G}\int d^4x\sqrt{|g|}\bigg\{R-2\Lambda&+7\lambda \tensor{R}{_{\mu\nu}^{\rho\sigma}}\tensor{R}{_{\rho\sigma}^{\alpha\beta}}\tensor{R}{_{\alpha\beta}^{\mu\nu}}  \\&  +7\tilde\lambda \tensor{R}{^{\mu}^{\nu}_{\alpha}_{\beta}}\tensor{R}{^{\alpha}^{\beta}_{\rho}_{\sigma}}\tensor{\tilde R}{^{\rho}^{\sigma}_{\mu}_{\nu}}+\mathcal{O}(\lambda^2,\tilde\lambda^2)\bigg\}\, .
\end{split}
\end{align}
The corrections to the thermodynamic properties of Kerr black holes in this theory (with $\Lambda=0$) at linear order in the couplings have been studied \cite{Reall:2019sah}. Thus, at linear order in $\lambda$ and $\tilde \lambda$, our results should match those of \cite{Reall:2019sah} with the identifications $\eta_{e}=7\lambda$, $\eta_{o}=7\tilde\lambda$.  However, the most interesting aspect of the theory \req{cubictheory} is that we do not need to restrict to the perturbative regime, as we shall be able to obtain an exact result. 

The equations of motion of \req{cubictheory} can be written as

\begin{equation}\label{eomcubic}
\mathcal{E}_{\mu\nu} = G_{\mu\nu}+\Lambda g_{\mu\nu}+\lambda \mathcal{E}^{\rm{even}}_{\mu\nu}+\tilde \lambda \mathcal{E}^{\rm{odd}}_{\mu\nu}=0\, ,
\end{equation}
where
\begin{align}
\mathcal{E}^{\rm{even}}_{\mu\nu}&=\tensor{R}{_{\mu}^{\sigma\alpha\beta}}P^{\rm even}_{\nu\sigma\alpha\beta}-\frac{1}{2}\mathcal{P}g_{\mu\nu}+2\nabla^{\alpha}\nabla^{\beta}P^{\rm even}_{\mu\alpha\nu\beta}\, ,\\
\mathcal{E}^{\rm{odd}}_{\mu\nu}&=\tensor{R}{_{\mu}^{\sigma\alpha\beta}}P^{\rm odd}_{\nu\sigma\alpha\beta}-\frac{1}{2}\tilde{\mathcal{P}}g_{\mu\nu}+2\nabla^{\alpha}\nabla^{\beta}P^{\rm odd}_{\mu\alpha\nu\beta}\, ,
\end{align}
and where

\begin{equation}
P_{\rm even}^{\mu\nu\alpha\beta}=\frac{\partial \mathcal{P}}{\partial R_{\mu\nu\alpha\beta}}\, ,\quad P_{\rm odd}^{\mu\nu\alpha\beta}=\frac{\partial \tilde{\mathcal{P}}}{\partial R_{\mu\nu\alpha\beta}}\, .
\end{equation}
These tensors read explicitly
\begin{align}\label{PECG}
    \begin{split}
		P_{\rm even}^{\beta_{1}\beta_{2}\beta_{3}\beta_{4}} &= 18R^{\beta_{2}\alpha\beta_{4}				\beta}R_{\alpha}{}^{\beta_{1}}{}_{\beta}{}^{\beta_{3}} - 18R^{\beta_{2}					\alpha\beta_{3}\beta}R_{\alpha}{}^{\beta_{1}}{}_{\beta}{}^{\beta_{4}} 
		+ 3R_{\alpha\beta}{}^{\beta_{1}\beta_{2}}R^{\beta_{3}\beta_{4}\alpha\beta}
		\\& \quad -6R^{\beta_{1}\beta_{3}}R^{\beta_{2}\beta_{4}} 6R^{\beta_{1}					\beta_{4}}R^{\beta_{2}\beta_{3}} + 12 R_{\nu\sigma} \left(R^{\nu\beta_{1}\sigma[\beta_{3}}				g^{\beta_{4}]\beta_{2}} - R^{\nu\beta_{2}\sigma[\beta_{4}}g^{\beta_{3}]\beta_{1}}\right)
		\\&\quad 
		+12R^{\beta_{4}\rho}R_{\rho}^{[\beta_{2}}g^{\beta_{1}]\beta_{3}} 
		- 12R^{\beta_{3}\rho}R_{\rho}^{[\beta_{2}}g^{\beta_{1}]\beta_{4}}\, ,
    \end{split}
\end{align}

\begin{align}\label{Pevil}
    \begin{split}
		P_{\rm odd}^{\beta_{1}\beta_{2}\beta_{3}\beta_{4}} 
		&=
		\frac{7}{2}\left( R^{\beta_{3}\beta_{4}}{}_{\rho\sigma}\tilde{R}^{\rho\sigma\beta_{1}			\beta_{2}} 
		+ R^{\beta_{1}\beta_{2}}{}_{\rho\sigma}\tilde{R}^{\rho\sigma\beta_{3}\beta_{4}}\right) 
		+ 7 R^{\beta_{3}\beta_{4}}{}_{\rho\sigma}\tilde{R}^{\beta_{1}\beta_{2}\rho\sigma} 
		+ 7 R^{\beta_{1}\beta_{2}}{}_{\rho\sigma}\tilde{R}^{\beta_{3}\beta_{4}\rho\sigma}
        \\&\quad
        -\frac{9}{2} g^{\beta_{1}[\beta_{3}}g^{\beta_{2}]\beta_{3}} \tensor{R}{_{\mu\nu\rho\sigma}}\tensor{\tilde{R}}{^{\mu\nu\rho\sigma}} -\frac{9}{2} R  \tilde{R}^{\beta_{1}\beta_{2}\beta_{3}\beta_{4}} -\frac{9}{2} R  \tilde{R}^{\beta_{3}\beta_{4}\beta_{1}\beta_{2}}
        \\& \quad
        -5 g^{-1/2}\epsilon^{\mu\nu\beta_{1}\beta_{2}}R^{[\beta_{3}}_{\mu}R^{\beta_{4}]}_{\nu} -5 g^{-1/2}\epsilon^{\mu\nu\beta_{3}\beta_{4}}R^{[\beta_{1}}_{\mu}R^{\beta_{2}]}_{\nu} \\& \quad
        -10 R_{\nu\sigma}\left(g^{\beta_{3}[\beta_{1}}\tilde{R}^{\beta_{2}]\nu\beta_{4}\sigma} -g^{\beta_{4}[\beta_{1}}\tilde{R}^{\beta_{2}]\nu\beta_{3}\sigma} + g^{\beta_{1}[\beta_{3}}\tilde{R}^{\beta_{4}]\nu\beta_{2}\sigma} -  g^{\beta_{2}[\beta_{3}}\tilde{R}^{\beta_{4}]\nu\beta_{1}\sigma} \right)\, ,
    \end{split}
\end{align}
and we must note that, in general, for both of these tensors, $P^{\mu[\nu\alpha\beta]}\neq 0$. As discussed in section~\ref{sec:charges}, the antisymmetric part is irrelevant for the equations of motion and for the entropy. However, in order to construct the Noether charge \req{eq: def of J} we will need to consider their ``barred'' versions $\bar{P}^{\mu\nu\alpha\beta}$ as in \req{eq: def of P}, \textit{i.e.}, with the antisymmetric part subtracted.

\subsection{Near-horizon geometries}
In order to study the near-horizon geometries of the theory \req{cubictheory}, it is convenient to use the ansatz \cite{Cano:2019ozf}
\begin{equation}\label{nhcubic}
ds^2_{N,f}=(a^2+x^2)\left(-r^2dt^2+\frac{dr^2}{r^2}\right)+\frac{dx^2}{f(x)}+N(x)^2f(x)\left(d\phi+2ar dt\right)^2\, ,
\end{equation}
which contains a parameter $a$ and two functions $N(x)$ and $f(x)$. The equations for these functions can be obtained equivalently from the equations of motion \req{eomcubic} or from the reduced Lagrangian

\begin{equation}
L[N,f]=N(x)(a^2+x^2)\mathcal{L}\big|_{\text{ansatz}}\, .
\end{equation}
In fact, the two cubic densities \req{ECG} and \req{EvilECG} satisfy the following property (also shared by the Einstein-Hilbert Lagrangian): when evaluated  on $N(x)=1$, the reduced Lagrangian $L[1,f]$ becomes a total derivative. In other words, this means that

\begin{equation}\label{GQTcondition}
\frac{\delta L}{\delta f}\Big|_{N=1}=0\, ,
\end{equation}
so the Euler-Lagrange equation for $f(x)$ is solved for $N(x)=1$. We can refer to this phenomenon by saying that the theories allow for ``single-function'' solutions, since we only need to determine the function $f(x)$.
All of the theories satisfying this property are a subset of the family of Generalized Quasi-topological Gravities (GQTG). Those are defined as the theories allowing for single-function static and spherically symmetric solutions \cite{Hennigar:2017ego,Bueno:2017sui} and they satisfy a similar condition to \req{GQTcondition} in that case. For a subset of GQTGs, this property is extended to solutions with NUT charge \cite{Bueno:2018uoy}.  
Indeed, one can see that the theories satisfying \req{GQTcondition} for near-horizon geometries are the same ones as the GQTGs allowing for single-function Taub-NUT solutions, which were originally studied in \cite{Bueno:2018uoy}. This is because one can transform the near-horizon ansatz into a Taub-NUT ansatz (in this case, with a hyperbolic base space) by performing a complex rotation $t\rightarrow i t$, $a\rightarrow i a$, where $\psi$ coordinate would play the role of Euclidean time coordinate.

It was already known that ECG \req{ECG} belonged to this subset of GQT theories. The density \req{EvilECG} is the first example of an odd-parity Lagrangian belonging to this class. While this density is trivial for spherically symmetric solutions, it contributes to the equations of motion for rotating near-horizon geometries (and equivalently for Taub-NUT geometries). 

Let us now discuss how to solve the equations of motion. The property \req{GQTcondition} means that the theory \req{cubictheory} allows for solutions with $N(x)=1$. At the level of the equations of motion, \req{GQTcondition} implies that
\begin{equation}
\mathcal{E}_{\psi\psi}\big|_{N=1}=f(x)^2\mathcal{E}_{xx}\big|_{N=1}\, ,
\end{equation}
so that, in fact, only one component of the equations of motion remains to be solved.\footnote{By using the Bianchi identities one can show that the rest of the components of the EOMs are proportional to $\mathcal{E}_{xx}$, $\mathcal{E}_{\psi\psi}$ and their derivatives, so there are no more independent equations.}
Furthermore, this equation takes the form of a total derivative, namely we have
\begin{equation}
\mathcal{E}_{\psi\psi}\big|_{N=1}=\frac{x^2 f(x)}{(a^2+x^2)}\frac{d}{dx}\mathcal{E}_{f}\, ,
\end{equation}
where 

\begin{equation}\label{feqcubic}
\begin{aligned}
\mathcal{E}_{f}&=x-\frac{a^2}{x}+\left(\frac{a^2}{x}+x\right) f+\Lambda  \left(-\frac{a^4}{x}+2 a^2 x+\frac{x^3}{3}\right)+\lambda\Bigg[-\frac{24 f^3 \left(a^4-9 a^2 x^2\right)}{x
	\left(a^2+x^2\right)^3}\\
&+\left(\frac{12 f^2 \left(-17
	a^2+x^2\right)}{\left(a^2+x^2\right)^2}+\frac{12 f}{a^2+x^2}\right)
f'+\left(\frac{6}{x}+\frac{12 a^2 f}{a^2 x+x^3}\right)
\left(f'\right)^2-2 \left(f'\right)^3\\
&+\left(-\frac{12
	f}{x}+\frac{f^2 \left(48 a^2-12 x^2\right)}{a^2 x+x^3}+6 f f'\right)
f''\Bigg]\\
&+\tilde \lambda\Bigg[-\frac{72 a f^2}{\left(a^2+x^2\right)^2}+\frac{8 f^3 \left(37 a^3-9 a
	x^2\right)}{\left(a^2+x^2\right)^3}+\left(\frac{84 f^2 \left(-a^3+a
	x^2\right)}{x \left(a^2+x^2\right)^2}+\frac{12 a f}{a^2
	x+x^3}\right) f'\\
&-\frac{12 a f \left(f'\right)^2}{a^2+x^2}-\frac{6 a
	\left(f'\right)^3}{x}+\left(-\frac{36 a f^2}{a^2+x^2}+\frac{18 a f
	f'}{x}\right) f''\Bigg]\, .
\end{aligned}
\end{equation}
Therefore, we have the equation 
\begin{equation}\label{feqcubicn}
\mathcal{E}_{f}=n\, ,
\end{equation}
where $n$ is an integration constant. Note that for $\lambda=\tilde \lambda=0$, this equation can be solved right away to obtain the solution in Einstein gravity,

\begin{align}
	N_0(x) = 1, \quad f_0(x)\ = \frac{a^2+n x-x^2+\Lambda\left(a^4  -2 a^2  x^2-x^4/3\right)}{\left(a^2+x^2\right)} \, .
\end{align}
This is the near-horizon geometry of the extremal Kerr-NUT-(A)dS solution, where $n$ is the NUT charge. Note that $n$ breaks the equatorial symmetry $x\leftrightarrow -x$ and as a consequence the horizon is non-smooth, as it contains a conical defect at least at one of the poles --- we will be more precise about this in a moment. 

Here we are interested only in solutions without NUT charge and hence with a smooth horizon, so in the case of Einstein gravity we must set $n=0$. The same applies for the even-parity cubic Lagrangian. However, the parity-breaking Lagrangian necessarily breaks the equatorial symmetry and we will find that a non-vanishing $n=n_{\rm smooth}$ is necessary in order to obtain a smooth horizon in that case. This should not be interpreted as a NUT charge, though; instead the NUT charge would be measured by the difference $n-n_{\rm smooth}$. 

\subsubsection*{Boundary conditions}
The equation \req{feqcubic} cannot be solved analytically when $\lambda, \tilde\lambda\neq 0$. One way to go around this consists in solving the equation perturbatively by performing an expansion in the higher-derivative couplings, analogous to what we did in section \ref{sec:heterotic} for the theory \req{Istringy}. However, it turns out that we can obtain exact information about the solution by studying the boundary conditions that need to be imposed. 

Since the solution must contain a compact horizon, the domain of $x$ must be compact. Thus, $f(x)$ must vanish at two different points
\begin{equation}\label{fx1x2}
f(x_1)=f(x_2)=0\, ,\quad x_2>x_1\, ,
\end{equation} 
which represent the north and south poles of the horizon. Furthermore, we want these poles to be smooth, \textit{i.e.}, free of conical defects. At this point we must take into account that $\phi$ is a periodic variable. Its period is a priori undetermined so let us write it as
\begin{equation}
	\phi \sim \phi+\frac{2\pi}{\omega}\, ,
\end{equation}
for a certain parameter $\omega>0$. Then one can check that smoothness at the poles is achieved if
\begin{equation}\label{fpx1x2}
f'(x_1)=2\omega\, ,\quad f'(x_2)=-2\omega\, .
\end{equation}
We additionally assume that the solution is analytic around the poles, and therefore it is given by series expansions of the form

\begin{equation}\label{fexp}
	f(x)=
\begin{cases}
	&\displaystyle+2\omega (x-x_1)+\sum_{k=2}^{\infty} a_k (x-x_1)^k\, ,\quad x\sim x_1\, ,\\
	&\displaystyle-2\omega (x-x_2)+\sum_{k=2}^{\infty }b_k (x-x_2)^k\, , \quad x\sim x_2\, ,
\end{cases}
\end{equation}
for certain coefficients $a_k$ and $b_k$.
We can now insert these expansions into the equation \req{feqcubicn} and solve order by order in $(x-x_i)^k$ in each case
\begin{equation}
\mathcal{E}_{f}=\sum_{k=0}^{\infty} \mathcal{E}_{f,i,k} (x-x_i)^k\, ,\quad i=1,2.
\end{equation}
This yields an infinite system of algebraic equations $\mathcal{E}_{f,i,k}=0$ involving the parameters of the solution and the coefficients $a_k$ and $b_k$. The first two equations, corresponding to $\mathcal{E}_{f,i,0}=0$ and $\mathcal{E}_{f,i,1}=0$, read as follows

\begin{equation}\label{constraineqs}
\begin{aligned}
x_1-\frac{a^2}{x_1}+\Lambda  \left(\frac{x_1^3}{3}+2 a^2 x_1-\frac{a^4}{x_1}\right)+\frac{8 \lambda  \omega ^2 \left(3-2 x_1 \omega
	\right)}{x_1}-\frac{48 a \tilde\lambda \omega ^3}{
	x_1}&=n\, ,\\
x_2-\frac{a^2}{x_2}+\Lambda  \left(\frac{x_2^3}{3}+2 a^2 x_2-\frac{a^4}{x_2}\right)+\frac{8 \lambda  \omega ^2 \left(3+2 x_2 \omega
	\right)}{x_2}+\frac{48 a \tilde\lambda \omega ^3}{x_2}&=n\, ,\\
\frac{\left(a^2+x_1^2\right) \left(1+2 x_1
	\omega \right)}{x_1^2}+\frac{\Lambda  \left(a^2+x_1^2\right)^2}{x_1^2}\\
+\frac{24 \lambda  \omega ^2 \left(x_1^2+a^2 \left(4 x_1 \omega
	-1\right)\right)}{x_1^2 \left(a^2+x_1^2\right)}+\frac{48 a
	\tilde\lambda \omega ^2 \left(a^2 \omega -x_1^2 \omega
	+x_1\right)}{x_1^2
	\left(a^2+x_1^2\right)}&=0\, ,\\
\frac{\left(a^2+x_2^2\right) \left(1-2 x_2
	\omega\right)}{x_2^2}+\frac{\Lambda  \left(a^2+x_2^2\right)^2}{x_2^2}\\
+\frac{24 \lambda  \omega ^2 \left(x_2^2-a^2 \left(4 x_2 \omega
	+1\right)\right)}{x_2^2 \left(a^2+x_2^2\right)}-\frac{48 a
	\tilde\lambda \omega ^2 \left(a^2 \omega-x_2^2\omega 
	-x_2\right)}{x_2^2
	\left(a^2+x_2^2\right)}&=0\, .
\end{aligned}
\end{equation}
These are four equations for the five parameters $x_{1}$, $x_{2}$, $n$, $\omega$ and $a$. Therefore, they determine four of the parameters in terms of the remaining one, for instance, $a$. Naturally, the remaining free parameter will be related to the angular momentum.
Note that in the case of $\tilde\lambda=0$ we have $n=0$ and $x_1=-x_2$, so that the problem is reduced to a system of two equations for three variables. 

A very remarkable property about these four equations is that they do not involve the parameters $a_k$ and $b_k$. These start appearing in $\mathcal{E}_{f,i,k}$ with $k\ge 2$ and when solving these equations one finds that the expansions in \req{fexp} are entirely determined by the first parameter, $a_2$ or $b_2$ (thus, there is only one integration constant in each case). These must be chosen so that the two solutions can be glued producing a complete solution which is regular at both poles. Such solution could in principle be obtained numerically via the shooting method, but however we will not need it for our purposes.   As we show next, we can obtain the entropy and the angular momentum even if we do not know the solution for $f(x)$ explicitly.

\subsection{Entropy and angular momentum}

\subsubsection*{Entropy}

In order to compute the black hole entropy we apply Iyer-Wald's formula \req{eq: IW entropy formula}, which for the theory \req{cubictheory} reads

\begin{equation}
S=\frac{1}{4G}\int d^2x\sqrt{h}\left[1-\frac{\lambda}{2}P_{\rm even}^{\mu\nu\alpha\beta}\epsilon_{\mu\nu}\epsilon_{\alpha\beta}-\frac{\tilde\lambda}{2}P_{\rm odd}^{\mu\nu\alpha\beta}\epsilon_{\mu\nu}\epsilon_{\alpha\beta}\right]\, ,
\end{equation}
where the tensors $P_{\rm even}^{\mu\nu\alpha\beta}$ and $P_{\rm odd}^{\mu\nu\alpha\beta}$ are those in \req{PECG} and \req{Pevil}. For the near-horizon geometry \req{nhcubic} with $N(x)=1$ the binormal has components
\begin{equation}
\epsilon_{tr}=-\epsilon_{rt}=a^2+x^2\, ,
\end{equation}
and the integral above becomes

\begin{equation}
S=\frac{\pi}{2\omega G}\int_{x_1}^{x_2} dx\left[1-2\lambda (a^2+x^2)^2 P_{\rm even}^{trtr}-2\tilde\lambda (a^2+x^2)^2 P_{\rm odd}^{trtr}\right]\, ,
\end{equation}
where we used the fact that $\sqrt{h}=1$ and integrated over the $\phi$ coordinate, which has periodicity $2\pi/\omega$. Remarkably, we find that this integrand takes the form of a total derivative. In particular, we have
\begin{align}
(a^2+x^2)^2 P_{\rm even}^{trtr}&=\frac{d}{dx}\left[\frac{3 x f'(x)^2}{a^2+x^2}+\frac{12 a^2 f(x) f'(x)}{\left(a^2+x^2\right)^2}-\frac{12 a^2 x f(x)^2}{\left(a^2+x^2\right)^3}\right]\, ,\\
(a^2+x^2)^2 P_{\rm odd}^{trtr}&=\frac{d}{dx}\left[\frac{3 a f'(x)^2}{ \left(a^2+x^2\right)}-\frac{12 a x f(x) f'(x)}{ \left(a^2+x^2\right)^2}-\frac{12 a^3 f(x)^2}{ \left(a^2+x^2\right)^3}\right]\,,
\end{align}
and therefore it is possible to carry out the integration explicitly even if we do not know what $f(x)$ is. Furthermore, using the boundary conditions \req{fx1x2} and \req{fpx1x2} we have the following result
\begin{align}\label{Scubic1}
S &= \frac{\pi}{2\omega G}\left[x_2-x_1+ 24\lambda  \omega^2  \left(\frac{ x_{1}}{a^2+x_{1}^2}-\frac{x_{2}}{a^2+x_{2}^2}\right)+  24\tilde{ \lambda}\omega^2  \left(\frac{a}{a^2+x_{1}^2}-\frac{ a}{a^2+x_{2}^2}\right)\right]\, .
\end{align}
In the limit that the $x_{2} \to -x_{1}$, we find that the contribution coming from the parity-violating term vanishes, as expected. This is an exact expression of the entropy in terms of the various parameters of the solution, but this is still not meaningful as our goal is to obtain the expression of $S$ in terms of the angular momentum $J$.

\subsubsection*{Angular momentum}
We compute the angular momentum from the Noether-Komar integral \req{eq: def of J}, where in this case 
\begin{equation}\label{Pbarcubic}
\bar{P}^{\mu\nu\rho\sigma}=\frac{1}{16\pi G}\left[g^{\mu[\rho} g^{\sigma] \nu}+\lambda\bar{P}_{\rm even}^{\mu\nu\rho\sigma}+\tilde\lambda \bar{P}_{\rm odd}^{\mu\nu\rho\sigma}\right]\, ,
\end{equation}
where we recall that the bar denotes that we subtract the antisymmetric part in the last three indices. The Killing vector we should consider is $\xi=\omega^{-1}\partial_{\phi}$, since $\omega \phi$ is the angular coordinate with $2\pi$ period. 
However, it will be useful to compute instead  the Komar charge for a general Killing vector
\begin{equation}\label{killingcubic}
\xi=\xi^{0}\partial_{t}+\xi^{3}\partial_{\phi}\, ,
\end{equation}
where $\xi^{0}$ and $\xi^{3}$ are constants. This will allow us to perform a consistency check of our results and will help us perform the integration for the angular momentum. 

The first step is to obtain the two-form $\boldsymbol{\Omega}$. For the metric \req{nhcubic} and the Killing vector \req{killingcubic}, we have

\begin{equation}
\diff\boldsymbol{\Omega}=\frac{1}{2}\left(\xi^{0} \diff r\wedge \diff \phi\wedge \diff x+\xi^{3}\diff t\wedge \diff r \wedge \diff x \right)(a^2+x^2)\mathcal{L}\, .
\end{equation}
Since the on-shell Lagrangian is only a function of $x$, an obvious solution of this equation is

\begin{equation}
\boldsymbol{\Omega}=-\frac{1}{2}\left(\xi^{0} \diff x\wedge \diff \phi+\xi^{3}\diff t  \wedge \diff x \right) r (a^2+x^2)\mathcal{L}\, .
\end{equation}
Now, for the same reasons discussed in section~\ref{sec:heterotic} below  \req{eq:omegaforstringy}, we find that this is the appropriate choice of $\boldsymbol{\Omega}$ in order to compute the angular momentum.

Let us also take note that, because the theory \req{cubictheory} belongs to the GQT class, the Lagrangian takes explicitly the form of a total derivative for any choice of $f(x)$. In fact, we have

\begin{equation}
(a^2+x^2)\mathcal{L}=\frac{1}{16\pi G}\frac{d\mathcal{I}}{dx}\, ,
\end{equation}
where

\begin{equation}
\begin{aligned}
\mathcal{I}(x)&=-\left(a^2+x^2\right) f'-2 f x-2 x-2 \Lambda  \left(a^2 x+\frac{x^3}{3}\right)+\lambda\Bigg[\frac{\left(10 a^2-2 x^2\right) \left(f'\right)^3}{a^2+x^2}\\
&+\left(f'\right)^2 \left(\frac{12 f x \left(x^2-9 a^2\right)}{\left(a^2+x^2\right)^2}+\frac{12 x}{a^2+x^2}\right)+f' \left(\frac{48 a^2 f}{\left(a^2+x^2\right)^2}-\frac{168 f^2 \left(a^4-a^2 x^2\right)}{\left(a^2+x^2\right)^3}\right)\\
&+\frac{48 f^3 \left(5 a^4 x-a^2 x^3\right)}{\left(a^2+x^2\right)^4}-\frac{48 a^2 f^2 x}{\left(a^2+x^2\right)^3}\Bigg]+\tilde\lambda \Bigg[-\frac{12 a x \left(f'\right)^3}{a^2+x^2}\\
&+\left(f'\right)^2 \left(\frac{f \left(48 a x^2-72 a^3\right)}{\left(a^2+x^2\right)^2}+\frac{12 a}{a^2+x^2}\right)+f' \left(\frac{48 f^2 \left(6 a^3 x-a x^3\right)}{\left(a^2+x^2\right)^3}-\frac{48 a f x}{\left(a^2+x^2\right)^2}\right)\\
&-\frac{48 a^3 f^2}{\left(a^2+x^2\right)^3}+\frac{32 f^3 \left(4 a^5-5 a^3 x^2\right)}{\left(a^2+x^2\right)^4}\Bigg]\, .
\end{aligned}
\end{equation}

The computation of the rest of the charge 2-form \req{eq: def of J} --- using the tensor \req{Pbarcubic} --- is straightforward. The full expression is lengthy and we show it in appendix~\ref{appendix: charge2form}. In order to obtain the conserved charge, we integrate it over a surface of constant $t$ and $r$, and hence we only need the $\diff x\wedge \diff \psi$ component. The result takes the form

\begin{equation}\label{Jint1}
J=\frac{1}{8\pi G}\int \diff x \diff \phi\left(r \xi^{0} F(x)+\xi^{3} G(x)\right)\,  ,
\end{equation}
where the functions $F(x)$ and $G(x)$ are given by \req{Ffunc} and \req{Gfunc}. Note that the term proportional to $\xi^{0}$ is also proportional to the radius of integration $r$. However, by construction, the Komar integral should be independent of the surface of integration. Thus, the only possibility is that this term gives zero contribution to the charge and its integral vanishes. In appendix~\ref{app:rintegral}, we show that this is indeed the case. The proof only requires using the equations of motion as well as the boundary conditions of the solution. As we already observed in section~\ref{subsec:SJheterotic}, this means that we cannot obtain the mass of the black hole from its near-horizon geometry. 

As the integral is independent of $r$, we are free to fix it to a value that facilitates the integration. In fact, we observe that, for  
\begin{equation}
r\xi^{0}=\frac{\xi^{3}}{2a}\, ,
\end{equation}
the integrand becomes an explicit total derivative.  We have

\begin{equation}
F(x)+\frac{1}{2a}G(x)=\frac{d \mathcal{J}(x)}{dx}\, ,
\end{equation}
where 

\begin{equation}
\begin{aligned}
 \mathcal{J}(x)&=-\frac{f x}{2 a}-\frac{\left(a^2+x^2\right) f'}{4 a}-\frac{\Lambda  x \left(3 a^2+x^2\right)}{6 a}-\lambda \Bigg[\frac{12 a f^3 x \left(-5 a^2+x^2\right)}{\left(a^2+x^2\right)^4}+\frac{42 a f^2
	\left(a^2-x^2\right) f'}{\left(a^2+x^2\right)^3}\\
&+\frac{3 f \left(9 a^2 x-x^3\right)
	\left(f'\right)^2}{a \left(a^2+x^2\right)^2}+\frac{\left(-5 a^2+x^2\right)
	\left(f'\right)^3}{2 a \left(a^2+x^2\right)}\Bigg]-\tilde\lambda\Bigg[-\frac{8 a^2 f^3 \left(4 a^2-5 x^2\right)}{\left(a^2+x^2\right)^4}\\
&+\frac{12 f^2 x \left(-6
	a^2+x^2\right) f'}{\left(a^2+x^2\right)^3}+\frac{6 f \left(3 a^2-2 x^2\right)
	\left(f'\right)^2}{\left(a^2+x^2\right)^2}+\frac{3 x \left(f'\right)^3}{a^2+x^2}\Bigg]\, .
\end{aligned}
\end{equation}
Therefore, by performing the integration in \req{Jint1}, we find that the angular momentum reads

\begin{equation}
J=\frac{1}{4\omega^2 G}\left[\mathcal{J}(x_2)-\mathcal{J}(x_1)\right]\, ,
\end{equation}
where we used that $\xi^{3}=1/\omega$ for the properly normalized rotational Killing vector. Finally, we can evaluate $\mathcal{J}(x_{1,2})$ by using the boundary conditions \req{fx1x2} and \req{fpx1x2}, and we find

\begin{equation}
\begin{aligned}\label{Jcubic}
J =& -\frac{\Lambda(x_2-x_1)(3a^2+x_1^2+x_1 x_2+x_2^2)}{24 a \omega^2 G}+\frac{2 a^2+x_{1}^2+x_{2}^2}{8 a G \omega }\\
&-\frac{2 \lambda  \omega  \left(5 a^4+2 a^2 \left(x_{1}^2+x_{2}^2\right)-x_{1}^2 x_{2}^2\right)}{a G \left(a^2+x_{1}^2\right) \left(a^2+x_{2}^2\right)}+\frac{6 \tilde{\lambda}\omega  (x_{1}+x_{2}) \left(a^2+x_{1} x_{2}\right)}{G \left(a^2+x_{1}^2\right) \left(a^2+x_{2}^2\right)}\, .
\end{aligned}
\end{equation}

\subsection{The relation $S(J)$}\label{subsec:relSJ}
By combining the result \req{Jcubic} together with the constraint equations \req{constraineqs}, all the constants $a$, $x_1$, $x_2$, $\omega$, $n$, and therefore, the full solution, is determined by the angular momentum --- up to the possible existence of several branches of solutions. In turn, using \req{Scubic1}, this defines implicitly the entropy as a function of the angular momentum $S(J)$. What is remarkable about our approach is that the results so far have involved no approximation whatsoever. The six equations \req{Jcubic}, \req{Scubic1} and \req{constraineqs} provide the exact result for the relationship $S(J)$ in the theory \req{cubictheory}. 
Of course, we still need to solve these equations if we want to obtain an explicit result. Let us therefore study several limits of interest.  

\subsubsection*{Asymptotically flat black holes}
Let us study first the case of $\Lambda=0$. 
The most relevant solutions correspond to those with large angular momentum, or more precisely, those with $J>> \lambda^{1/2}/G$. This is the regime where the higher-derivative corrections are perturbatively small and the theory \req{cubictheory} makes sense as an effective field theory. The result in this case can be expressed as a power series in $1/J$ (equivalently as a power series in $\lambda$ and $\tilde\lambda$) and the few first terms read

\begin{align}\label{Scubicpert}
	S &= 2 \pi  |J|\left[1-\frac{\lambda }{(GJ)^2}+\frac{20 \lambda ^2+9 \tilde{\lambda}^2}{2 (G J)^4}-\frac{184 \lambda^3+117 \lambda  \tilde\lambda^2}{2 (G J)^6}+O\left(J^{-8}\right)\right]\, .
\end{align}
Importantly, the first order correction $\mathcal{O}(\lambda)$ agrees with the result by Reall and Santos \cite{Reall:2019sah} once we notice that $\lambda=\eta_{e}/7$. Our method is able to compute the subleading corrections as well, and in particular, we observe that the parity-breaking density does affect the entropy starting at order $\mathcal{O}(\tilde\lambda^2)$. 

\begin{figure}[t!]
	\centering
	\includegraphics[width=0.48\textwidth]{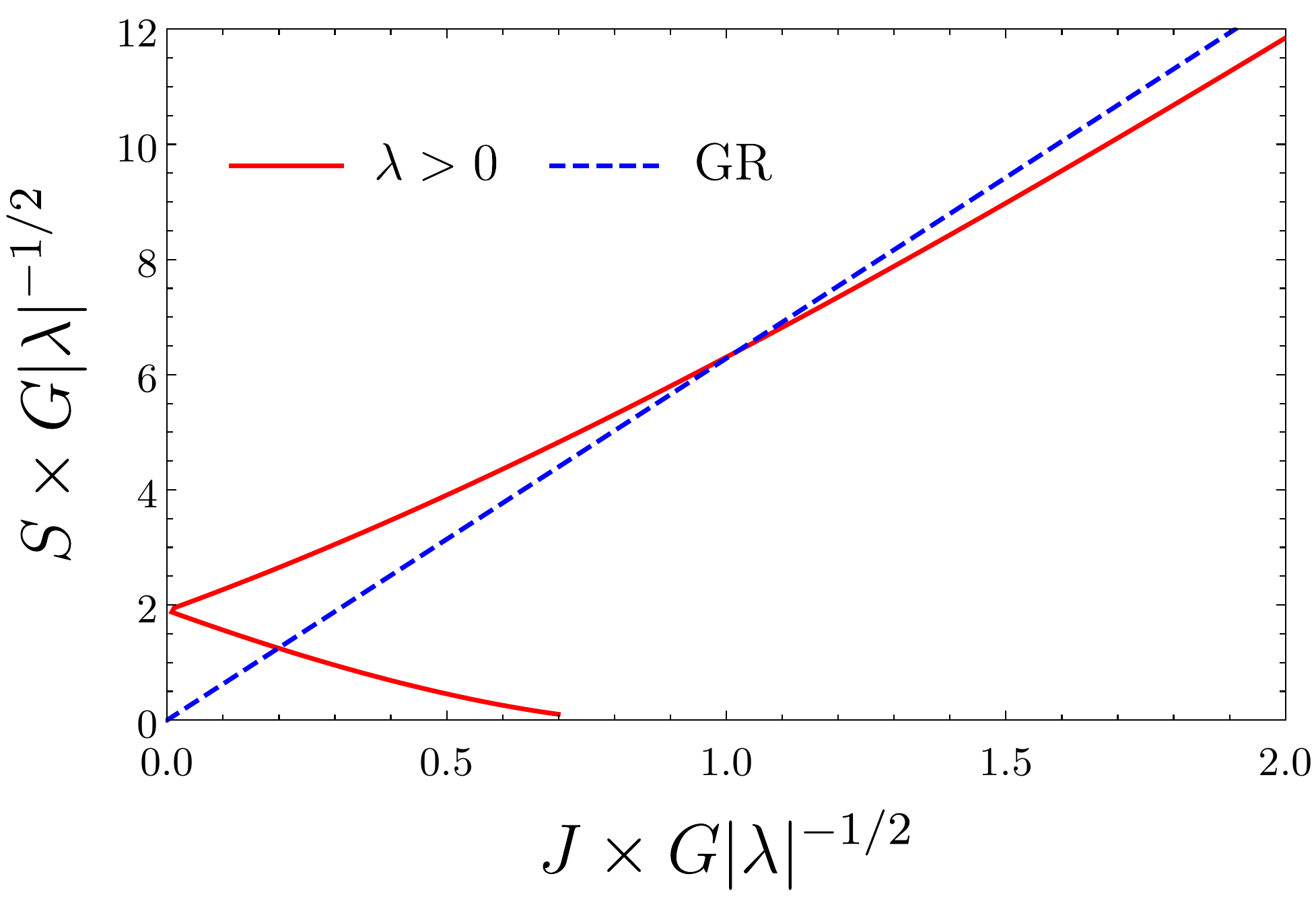}
	\includegraphics[width=0.48\textwidth]{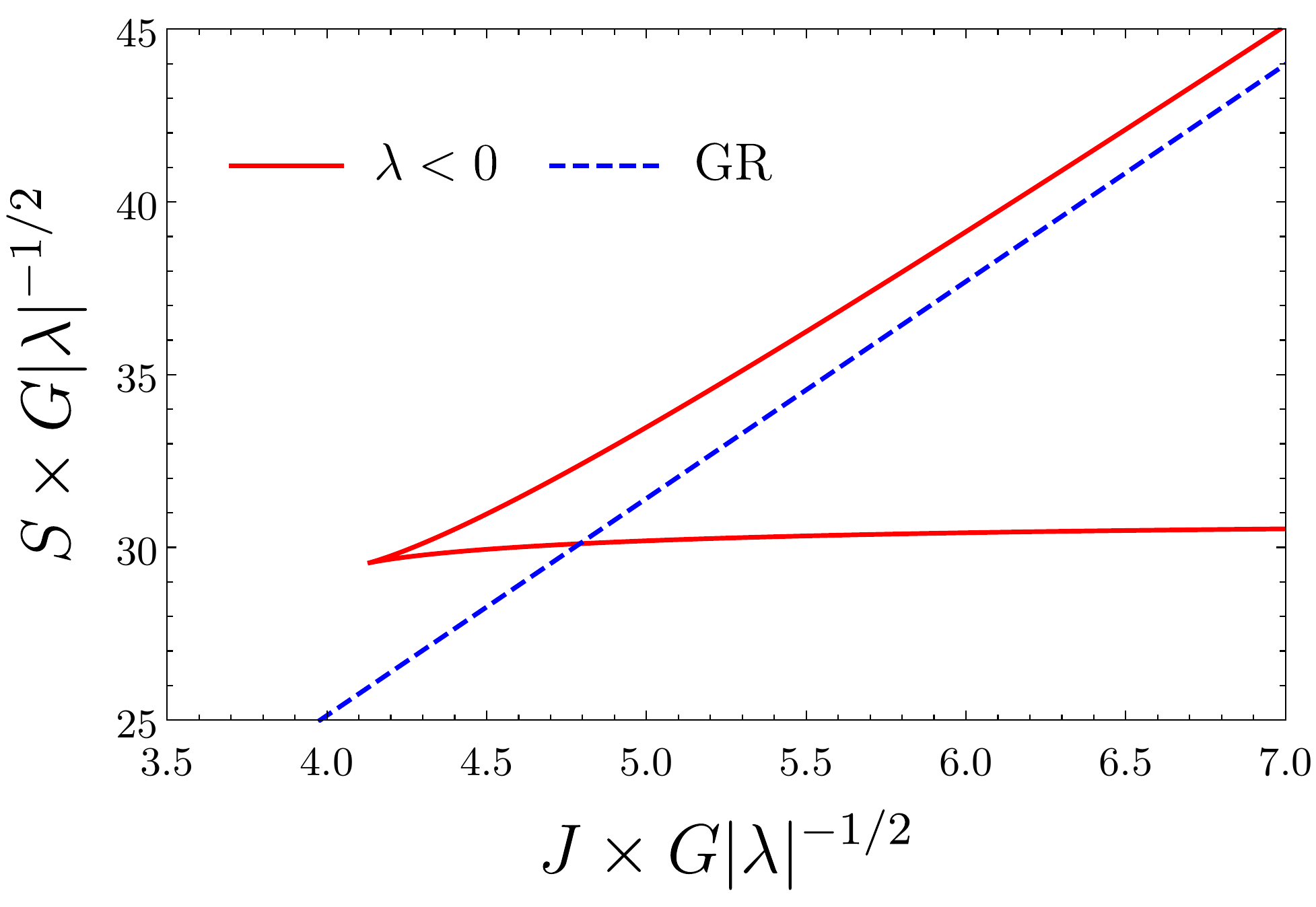}
	\caption{The relation $S(J)$ in the presence of the even-parity cubic correction. Left: $\lambda>0$, right: $\lambda<0$. The GR prediction $S=2\pi |J|$ is shown for comparison.}
	\label{SvsJflat}
\end{figure}

We can go beyond the perturbative result by numerically solving the system of equations \req{constraineqs} for finite values of the couplings. For simplicity, we focus on the parity-preserving correction so we set $\tilde\lambda=0$. Also, since there is only one scale in the problem, we can introduce the entropy and angular momentum in ``natural units'' $\hat S=S\times G|\lambda|^{-1/2}$ and $\hat J=J G|\lambda|^{-1/2}$, so that the relation $\hat{S}(\hat J)$ is independent of $|\lambda|$ and we only need to distinguish between $\lambda>0$ and $\lambda<0$. The result is shown in Figure~\ref{SvsJflat}, where we also include the GR result $S=2\pi|J|$ for comparison. 

In these plots we can observe some genuine non-perturbative effects. For $\lambda>0$ we see that at $J=0$ there is a solution with non-vanishing entropy (and also non-vanishing area). Also for small values of $J$ there is a second branch of solutions that ends at a solution with $S=0$ and $J\neq 0$. On the other hand, for $\lambda<0$ we see that there are no solutions below a minimum value of the angular momentum, and we have approximately the bound 
\begin{equation}
|J|>4.1 \frac{\sqrt{|\lambda|}}{G}\, .
\end{equation}
Above this value there are two solutions: one that asymptotes to Kerr for large $J$ and whose entropy behaves as \req{Scubicpert}, and a second one whose entropy tends to a constant value and which does not exist in the GR limit.
It is not clear what credibility one can assign to these non-perturbative phenomena since they lie beyond the limit of validity of effective field theory, but they provide an interesting example of the effects of higher-derivative corrections in a highly non-linear regime.

\subsubsection*{Adding a cosmological constant}
Let us now consider $\Lambda\neq 0$. The cosmological constant introduces a new scale and it gives rise to a non-linear relation $S(J)$ even for Einstein gravity, so let us start by reviewing that case.

When $\lambda=\tilde\lambda=0$ we have $x_1=-x_2$ and $n=0$. Then, introducing the dimensionless variable, 
\begin{equation}
\zeta=\frac{x_2}{a}\, ,
\end{equation}
we can solve the equations \req{constraineqs} as
\begin{equation}\label{aomegasol}
a=\sqrt{\frac{3(1-\zeta ^2)}{\Lambda ( \zeta^4+6\zeta^2-3)}}\, ,\quad  \omega=\frac{2 \zeta \left(3-\zeta^2\right)}{(\zeta^4+6 \zeta^2-3)a}.
\end{equation}
Observe that demanding that $a$ is real and that $\omega>0$ leads to the following bounds on $\zeta$ depending on the sign of $\Lambda$:\footnote{In the dS case one may also consider the solutions with $0<\zeta< \sqrt{2\sqrt{3}-3}$. Interestingly, these do not correspond to near-horizon geometries but to rotating Nariai geometries \cite{1950SRToh..34..160N} (consistent of a fibration of dS$_{2}$ over $\mathbb{S}^2$) which arise when the black hole horizon merges with the cosmological horizon.} 

\begin{equation}
\begin{aligned}
\Lambda>0\quad  &\Rightarrow \quad \sqrt{2\sqrt{3}-3}\le \zeta \le 1\, ,\\
\Lambda<0\quad  &\Rightarrow \quad 1\le \zeta\le \sqrt{3}\, .
\end{aligned}
\end{equation}
  
Now, plugging this into our expressions for the entropy \req{Scubic1} and the angular momentum \req{Jcubic} we get 
\begin{align}\label{SLambda0}
S_0=&\frac{3 \pi  \left(1-\zeta^2\right)}{G \Lambda  \left(3-\zeta^2\right)}\, ,\\
J_0=&\frac{3 \zeta  \left(1-\zeta^2\right)}{G \Lambda  \left(3-\zeta^2\right)^2}\, ,
\label{JLambda0}
\end{align}
which are precisely the relations for the Kerr-(A)dS black hole that we obtained in \req{eq: J with Lambda leading term heterotic}. The explicit expression for $S(J)$ is very involved because it requires solving a fourth degree polynomial, so it is more useful to work with this parametric relation.

Having obtained the result for every quantity $a$, $x_{i}$, $\omega$ and $n$, for Einstein gravity, we can now solve the equations \req{constraineqs} in the case of $\lambda\neq0$, $\tilde\lambda\neq0$ by assuming a series expansion around that result,

\begin{equation}
a=\sum_{p,q} a_{(p,q)}\lambda^{p}\tilde\lambda^{q}\, ,\quad x_{i}=\sum_{p,q} x_{i\,(p,q)}\lambda^{p}\tilde\lambda^{q}\, ,\quad \omega=\sum_{p,q} \omega_{(p,q)}\lambda^{p}\tilde\lambda^{q}\, ,\quad n=\sum_{p,q} n_{(p,q)}\lambda^{p}\tilde\lambda^{q}\, ,
\end{equation}
where $n_{(0,0)}=0$, $x_{2(0,0)}=-x_{1(0,0)}=a_{(0,0)}\zeta$ and $a_{(0,0)}$ and $\omega_{(0,0)}$ are those given by \req{aomegasol}.
At every order in $\lambda^{p}\tilde\lambda^{q}$ we obtain a linear system of equations for the coefficients $a_{(p,q)}$, $x_{i\, (p,q)}$, $\omega_{(p,q)}$ and $n_{(p,q)}$ that we can solve straightforwardly. Plugging back the result in \req{Scubic1} and \req{Jcubic} we then obtain the corrections to \req{SLambda0} and \req{JLambda0} expressed as a function of $\zeta$. However, since $\zeta$ is an arbitrary parameter, we can consider a redefinition 
\begin{equation}
\zeta\rightarrow \zeta+\sum_{p,q} \delta\zeta_{(p,q)}\lambda^p\tilde\lambda^q =\zeta+\delta\zeta_{(1,0)}\lambda+\delta\zeta_{(0,1)}\tilde\lambda+\ldots \, .
\end{equation} 
By choosing the coefficients $\delta\zeta_{(p,q)}$ appropriately we can always make the expression of $J$ as a function of $\zeta$ identical to that of Einstein gravity \req{JLambda0}. Thus, with the appropriate choice of $\zeta$, we have $J=J_0$. For that choice, the entropy reads, to quadratic order in $\lambda$ and $\tilde\lambda$

\begin{equation}\label{eq:ScubicLambda}
\begin{aligned}
S=S_0&\left[1-\frac{16  \lambda  \Lambda ^2\zeta ^2 \left(\zeta ^2-3\right)^2 \left(\zeta ^4+13 \zeta ^2-12\right)}{9 \left(\zeta ^2-1\right)^2 \left(\zeta ^6+7 \zeta ^4+3 \zeta ^2-3\right)}+\frac{64 \zeta ^2 \left(\zeta ^2-3\right)^4\lambda ^2 \Lambda ^4}{81 \left(\zeta ^4-1\right)^4 \left(\zeta ^4+6 \zeta ^2-3\right)^3} \left(\zeta ^{18}+70 \zeta ^{16}\right.\right.\\
&\left.\left.-582 \zeta ^{14}-746 \zeta ^{12}+13444 \zeta ^{10}+2946 \zeta ^8-34650 \zeta ^6+33282 \zeta ^4-13797 \zeta ^2+2592\right)\right.\\
&\left.+\frac{128 \tilde \lambda^2\Lambda ^4 \zeta^2 \left(3-5 \zeta ^2\right)^2 \left(\zeta ^2-3\right)^4 }{9 \left(\zeta ^4-1\right)^4 \left(\zeta ^4+6 \zeta ^2-3\right)} +\ldots \right]\, .
\end{aligned}
\end{equation}
An explicit expression $S(J)$ can be obtained in the limits of small and large black holes. For small black holes, $J<<1/(G|\Lambda|)$, which are obtained in the limit $\zeta\rightarrow 1$, the effect of the cosmological constant is irrelevant and we reproduce the asymptotically flat result \req{Scubicpert}. Large black holes, \textit{i.e.}, those with $J>>1/(G|\Lambda|)$,  are only possible in the AdS case $\Lambda<0$, and in that limit (corresponding to $\zeta\rightarrow \sqrt{3}$) we have

\begin{equation}
\begin{aligned}
S(J)=S_0(J)\left[1+\lambda\left(\frac{3 \sqrt{3} \Lambda }{G J}-\frac{2^{3/2}3^{1/4}\sqrt{-\Lambda }}{(GJ)^{3/2}}+\frac{1}{4 (G J)^2}\right)+\frac{27 (\lambda ^2+\tilde\lambda^2) \Lambda ^2}{4 (G J)^2}+\mathcal{O}\left(J^{-5/2}\right)\right]\, .
\end{aligned}
\end{equation} 
Unlike the case of the stringy corrections for large AdS black holes \req{SstringylargeJ}, the correction due to the cubic terms vanishes for $J\rightarrow\infty$.


\section{Discussion}
\label{sec:conclusions}
In this paper we have computed the corrections to the entropy-angular momentum relation of the extremal Kerr black hole in two different higher-derivative extensions of GR  motivated by string theory \req{Istringy} and by general effective field theory arguments \req{Icubic}. 
Our strategy consisted in studying near-horizon extremal geometries, since the existence of an enhanced symmetry in the near-horizon region greatly simplifies the analysis of the solutions. One can directly obtain the black hole entropy from the near-horizon metric thanks to the Iyer-Wald formula. 
However, one should express the entropy in terms of physically meaningful quantities of the black hole --- in this case, the angular momentum --- rather than some arbitrary parameters of the solution.
Computing the angular momentum from the near-horizon geometry is more challenging, since we lack an asymptotic region in our solutions. In fact, the angular momentum must identified as the integral of the Noether charge two-form in a sphere at infinity. The integration surface cannot be deformed as, in general, the charge two-form does not satisfy a Gauss law. In order to address this problem, we provided an appropriate Komar-type modification of the charge that is independent of the surface of integration. 
The Komar generalization of the Noether charge two-form is in general ambiguous, but, for the type of solutions considered, we demonstrated that there is a precise choice of the Noether-Komar two-form whose integral on the black hole horizon computes the angular momentum unambiguously.

Another method to identify the angular momentum from the near-horizon geometry was proposed in \cite{Astefanesei:2006dd} in the context of the entropy function approach. 
The underlying idea of \cite{Astefanesei:2006dd} is that the angular momentum can be identified as an electric charge in one dimension less, after compactification of the spacetime in the periodic coordinate. It is not evident that this method is equivalent to the Noether charge approach we employed. However, we checked that for the solutions we studied, both methods yield the same answer.

Besides near-horizon geometries, the only other approach we are aware of that can be used to study the corrections to the entropy of extremal rotating black holes is that of Reall and Santos \cite{Reall:2019sah}. This approach has the advantage of being applicable to non-extremal black holes, but has the limitation of only working for first-order corrections. The near-horizon approach on the other hand allows us to go beyond the leading correction. 
In our case, we were able to obtain the corrections to the entropy at order $\alpha'^2$ for the stringy action \req{Istringy} while for the cubic theory \req{Icubic} we obtained the exact result --- implicitly defined as the solution of a system of algebraic equations. 

On the other hand, the near-horizon geometry does not allow us to obtain the mass of the solutions. For first-order corrections, the mass could be obtained using the methods of \cite{Reall:2019sah} or \cite{Aalsma:2021qga}, but beyond first order the computation of the full global solution seems unavoidable in order to obtain the mass. 

Let us now discuss the specific results we obtained. In the case of the heterotic string theory, we were able to find explicitly the corrected NHEK geometry. This is even more remarkable in the Kerr-(A)dS case, where the solution takes a quite involved, yet fully analytic form. 
Obtaining the solution was necessary in order to compute both the entropy and the angular momentum, which allowed us to derive the relation \req{eq: entropy for heterotic string intro} (and \req{eq: S with Lambda zeta heterotic} for the case of $\Lambda\neq 0$). 
While the $\mathcal{O}(\alpha')$ correction to the entropy --- entirely coming from the topological Gauss-Bonnet term --- is positive, we observe that the $\mathcal{O}(\alpha'^2)$ correction is negative. 
It would be interesting to understand how this is related to the fact that we are assuming a stabilized dilaton. In the case where the dilaton remains massless, the corrections to the entropy of the non-extremal Kerr black hole were shown to be positive \cite{Cano:2021rey}. The extremal solution becomes singular in that case --- which is the reason why we consider a stabilized dilaton --- but studying the near-extremal solutions could shed light on the nature of this singularity.\footnote{As recently shown by \cite{Horowitz:2023xyl}, singularities are in fact very common in extremal rotating black holes with higher-derivative corrections. Even in the cases where the background geometry remains regular, fluctuations around it may render the spacetime singular. In this sense, the singularity due to the dilaton is perhaps not so surprising.} It would be interesting to find out whether there is a well-defined limit for the entropy at extremality when the dilaton is massless.

We also obtained the entropy of extremal rotating black holes for the cubic theories \req{Icubic}, that provide a general effective field theory extension of GR to six derivatives. These cubic Lagrangians are chosen in a special way: they correspond to Einsteinian cubic gravity \cite{Bueno:2016xff} and to a new parity-violating density with analogous properties to ECG that we introduced here for the first time.
The most remarkable property of these theories is that certain aspects of the near-horizon geometries, like the entropy and the angular momentum, can be obtained exactly even if we do not know the solution explicitly. Indeed, the full non-perturbative relation $S(J)$ is obtained from the equations \req{Scubic1}, \req{Jcubic} and \req{constraineqs}. This allows us to easily compute the perturbative expansion of $S(J)$ at virtually any order and even to study it in a non-perturbative fashion, as discussed in section~\ref{subsec:relSJ}. 
The fact that we can obtain the pertubative expansion at higher orders is highly non-trivial and relevant from the point of view of effective field theory. Let us remember that the effective action for gravity contains an infinite tower of higher-derivative corrections. However, each term in the action generates at the same time an infinite number of corrections on the black hole entropy. Here we have been able to provide the complete answer for the corrections to the entropy related exclusively with the six-derivative terms. 
Higher-derivative terms will induce additional corrections that should be added to this result. For instance, the $\mathcal{O}(\lambda^2,\tilde\lambda^2)$ terms in \req{eq: S cubic} contribute to the entropy at the same order as the leading contribution from 10-derivative terms. 
Our results have also showed for the first time the effect of a parity-breaking correction on the entropy. Interestingly the sign of this correction is fixed since it enters at order $\tilde\lambda^2$, and it is always positive. 

In sum, the strategy of writing the effective field theory of gravity in terms of these special densities --- which belong to the broader family of Generalized Quasitopological gravities \cite{Hennigar:2017ego,Bueno:2017sui,Bueno:2018uoy} --- greatly simplifies the study of solutions and allow us to go beyond the leading-order corrections. It would be interesting to explore if this procedure can be generalized to higher-order terms (beyond six derivatives) \cite{Bueno:2019ltp}.

Finally, this work could be extended in several ways.
Here we restricted ourselves to the case of pure gravity, but it would be interesting to study charged solutions as well. On the one hand, one may consider a general effective field theory extension of Einstein-Maxwell theory and study the corrections to the extremal Kerr-Newman solution. On the other, one could consider the effective action of heterotic string theory with gauge fields and study the corrections to the Kerr-Sen solution \cite{Sen:1992ua}. 
Other cases with additional fields, higher dimensions and/or supersymmetry --- like the BMPV black hole \cite{Breckenridge:1996is,Guica:2005ig} --- would also be worth to be explored.


\section*{Acknowledgments}
We thank Nikolay Bobev, Evan McDonough, Rishi Mouland, David Pere\~niguez, Thomas Van Riet and Nicolas Yunes for useful discussions. The work of PAC is supported by a postdoctoral fellowship from the Research Foundation - Flanders (FWO grant 12ZH121N). MD is supported by KU Leuven C1 grant ZKD1118 C16/16/005, and by the Research Programme of The Research Foundation – Flanders (FWO) grant G0F9516N.

\appendix

\section{Near-horizon geometry of the heterotic theory with $\Lambda\neq 0$} \label{appendix: heterotic string}
The first nonzero $\alpha'$ corrections for $\psi, f,$ and $g$ in the presence of a nonzero cosmological constant were found in section~\ref{sec:heterotic}. The various parameters in \eqref{eq: psi1 with lambda heterotic}, \eqref{eq: g2 with lambda heterotic}, and \eqref{eq: f2 with lambda heterotic} are given by
\begin{align}
	\begin{split}
		\rho_{f,0} &= -y_{0}^4-3,
		\\
		\rho_{f,1} &= -4 y^2 y_{0}^2 \left(y_{0}^2+3\right),
		\\
		\rho_{f,2} &= y^4 \left(y_{0}^4-4 y_{0}^2+7\right),
	\end{split}
	\\ \nonumber \\
	\begin{split}
		\kappa_{f,0} &= 7 y_{0}^{10}+48 y_{0}^8+96 y_{0}^6-18 y_{0}^4-135 y_{0}^2+162,
		\\
		\kappa_{f,2} &= y_{0}^{10}+15 y_{0}^8+78 y_{0}^6-6 y_{0}^4-255 y_{0}^2+135,
		\\
		\kappa_{f,4} &= -3 \left(y_{0}^8+2 y_{0}^6-8 y_{0}^4+14 y_{0}^2-9\right),
	\end{split}
	\\ \nonumber \\
	\begin{split}
		\tau_{f,0} &= -47 y_{0}^{12}-286 y_{0}^{10}-683 y_{0}^8-1124 y_{0}^6-685 y_{0}^4+354 y_{0}^2-281,
		\\
		\tau_{f,2} &= -7 y_{0}^{12}-70 y_{0}^{10}-273 y_{0}^8-420 y_{0}^6+343 y_{0}^4+522 y_{0}^2-287,
		\\
		\tau_{f,4} &= 2 \left(10 y_{0}^{10}+19 y_{0}^8-4 y_{0}^6+74 y_{0}^4-102 y_{0}^2+3\right),
	\end{split}
	\\ \nonumber \\
	\begin{split}
\gamma_{g,0} &= -4 \left(88 y_{0}^{10}+644 y_{0}^8+3271 y_{0}^6+9237 y_{0}^4-14403 y_{0}^2-765\right),
\\
\gamma_{g,2} &= 
-4873 y_{0}^{10}-23429 y_{0}^8-76486 y_{0}^6-224022 y_{0}^4+209343 y_{0}^2-68085,
\\
\gamma_{g,4} &=60 \left(246 y_{0}^{10}+2093 y_{0}^8+4862 y_{0}^6+1772 y_{0}^4+12452 y_{0}^2-2433\right),
\\
\gamma_{g,6} &=14 \left(1093 y_{0}^{10}+8435 y_{0}^8+18310 y_{0}^6+7794 y_{0}^4+58053 y_{0}^2-13653\right),
\\
\gamma_{g,8} &=140 \left(64 y_{0}^{10}+437 y_{0}^8+883 y_{0}^6+543 y_{0}^4+3273 y_{0}^2-792\right),
\\
\gamma_{g,10} &=105 \left(19 y_{0}^{10}+123 y_{0}^8+242 y_{0}^6+186 y_{0}^4+939 y_{0}^2-261\right),
\end{split}
\\ \nonumber \\
\begin{split}
\gamma_{f,0} &= 4 \left(298 y_{0}^{14}+2324 y_{0}^{12}+5635 y_{0}^{10}+13689 y_{0}^8+27540 y_{0}^6+4266 y_{0}^4+24831 y_{0}^2-2295\right),
\\
\gamma_{f,2} &=
3 \left(636 y_{0}^{14}+5741 y_{0}^{12}+14566 y_{0}^{10}+74875 y_{0}^8+96936 y_{0}^6-213333 y_{0}^4+130550 y_{0}^2 \right. \\& \left. \quad\quad -27795\right),
\\
\gamma_{f,4} &=
2 \left(27574 y_{0}^{14}+257531 y_{0}^{12}+966888 y_{0}^{10}+2159253 y_{0}^8+2287958 y_{0}^6-53427 y_{0}^4\right. \\& \left. \quad\quad+966092 y_{0}^2-50205\right),
\\
\gamma_{f,6} &=
37120 y_{0}^{14}+378777 y_{0}^{12}+1529990 y_{0}^{10}+3881051 y_{0}^8+3496412 y_{0}^6-3369561 y_{0}^4\\&  \quad\quad+2335518 y_{0}^2-345627,
\\
\gamma_{f,8} &=
12 \left(2198 y_{0}^{14}+23291 y_{0}^{12}+95919 y_{0}^{10}+254192 y_{0}^8+221264 y_{0}^6-310209 y_{0}^4+217323 y_{0}^2\right. \\& \left. \quad\quad-35370\right),
\\
\gamma_{f,10} &=
7 \left(980 y_{0}^{14}+10879 y_{0}^{12}+44586 y_{0}^{10}+142065 y_{0}^8+126952 y_{0}^6-332583 y_{0}^4+256282 y_{0}^2\right. \\& \left. \quad\quad-48201\right),
\\
\gamma_{f,12} &=
70 \left(6 y_{0}^{14}+25 y_{0}^{12}-56 y_{0}^{10}+1187 y_{0}^8+2006 y_{0}^6-10041 y_{0}^4+10140 y_{0}^2-2115\right),
\\
\gamma_{f,14} &=
-105 \left(17 y_{0}^{12}+110 y_{0}^{10}+139 y_{0}^8-116 y_{0}^6+735 y_{0}^4-1146 y_{0}^2+261\right).
\end{split}
\end{align}

\section{Noether-Komar two-form for the cubic theories}\label{appendix: charge2form}

The Noether-Komar 2-form \req{Komar2form} has the following components

\begin{equation}
\begin{aligned}
\tilde{\boldsymbol{Q}}_{\xi}=\frac{1}{16\pi G}\Big[&\diff t\wedge \diff r \left(A r \xi^{0}+B\xi^3\right)+ \diff t\wedge \diff x\left(C r^2 \xi^{0}+D r\xi^3\right)+\diff r\wedge \diff \phi \xi^0 E\\
&+\diff x\wedge \diff \phi \left(F r \xi^{0}+G\xi^3\right)\Big]\, ,
\end{aligned}
\end{equation}
where $A$, $B$, $C$, $D$, $E$, $F$ and $G$ are the following functions. In these expressions we use the dimensionless variable $y=x/a$ and the derivatives are with respect to $y$, \textit{e.g.}, $f'=\frac{df}{dy}=a\frac{df}{dx}$. 

\begin{align}\notag
A=&a^2 \left(-2 f y+\left(1+y^2\right) f'\right)+\frac{\lambda}{a^2} 
\Bigg[\frac{8 f^2 y \left(27-5 y^2\right)}{\left(1+y^2\right)^3}-\frac{40 f
	y}{\left(1+y^2\right)^2}-\frac{32 f^3 y \left(-35+58
	y^2\right)}{\left(1+y^2\right)^4}\\\notag
&+\left(\frac{20}{1+y^2}+\frac{24 f \left(-9+2
	y^2\right)}{\left(1+y^2\right)^2}-\frac{4 f^2 \left(29-576 y^2+15
	y^4\right)}{\left(1+y^2\right)^3}\right) f'\\\notag
&+\left(-\frac{68 y}{1+y^2}+\frac{8 f y
	\left(-50+3 y^2\right)}{\left(1+y^2\right)^2}\right) \left(f'\right)^2+\frac{6
	\left(-5+y^2\right) \left(f'\right)^3}{1+y^2}+\left(\frac{40 f y}{1+y^2}\right.\\\notag
&\left.+\frac{f^2
	\left(-676 y+60 y^3\right)}{\left(1+y^2\right)^2}+\left(10+\frac{f \left(206-42
	y^2\right)}{1+y^2}\right) f'\right) f''+6 f y \left(f''\right)^2\\\notag
&+\left(-12 f-\frac{12 f^2
	\left(-4+y^2\right)}{1+y^2}+6 f y f'\right) f^{(3)}\Bigg]+\frac{ \tilde\lambda}{a^2} \Bigg[-\frac{24
	f}{\left(1+y^2\right)^2}+\frac{8 f^2 \left(23+39 y^2\right)}{\left(1+y^2\right)^3}\\\notag
&+\frac{8
	f^3 \left(-11-432 y^2+63 y^4\right)}{\left(1+y^2\right)^4}+\left(-\frac{12
	y}{1+y^2}+\frac{f^2 \left(1952 y-720 y^3\right)}{\left(1+y^2\right)^3}+\frac{f \left(76 y-12
	y^3\right)}{\left(1+y^2\right)^2}\right) f'\\\notag
&+\left(\frac{2 \left(-33+5
	y^2\right)}{1+y^2}+\frac{4 f \left(-21+53 y^2\right)}{\left(1+y^2\right)^2}\right)
\left(f'\right)^2+\frac{36 y \left(f'\right)^3}{1+y^2}+\left(4+\frac{4 f
	\left(5+y^2\right)}{1+y^2}\right.\\
&\left.+\frac{8 f^2 \left(-39+37
	y^2\right)}{\left(1+y^2\right)^2}+\left(-6 y-\frac{216 f y}{1+y^2}\right) f'\right) f''+18 f
\left(f''\right)^2+\left(-\frac{36 f^2 y}{1+y^2}+18 f f'\right) f^{(3)}\Bigg]\, ,
\end{align}

\begin{align}\notag
B&=\frac{1}{2} a \left(-1-y^2\right) f'+\frac{\lambda}{a^3} \Bigg[\frac{8 f^2
	y}{\left(1+y^2\right)^3}+\frac{16 f^3 y \left(-1+14
	y^2\right)}{\left(1+y^2\right)^4}+\left(\frac{4 f \left(8+3
	y^2\right)}{\left(1+y^2\right)^2}\dvv+\frac{4 f^2 \left(-31-42 y^2+3
	y^4\right)}{\left(1+y^2\right)^3}\right) f'+\left(\frac{12 y}{1+y^2}+\frac{f \left(-62 y+6
	y^3\right)}{\left(1+y^2\right)^2}\right) \left(f'\right)^2-\frac{3 \left(-5+y^2\right)
	\left(f'\right)^3}{1+y^2}\dv+\left(-\frac{12 f y}{1+y^2}+\frac{f^2 \left(88 y-12
	y^3\right)}{\left(1+y^2\right)^2}+\frac{f \left(-20+6 y^2\right) f'}{1+y^2}\right)
f''\Bigg]+\frac{\tilde\lambda}{a^3}\Bigg[-\frac{24 f^2 \left(2+3
	y^2\right)}{\left(1+y^2\right)^3}+\dv \frac{8 f^3 \left(20+45 y^2-9
	y^4\right)}{\left(1+y^2\right)^4}+\left(-\frac{12 f y}{\left(1+y^2\right)^2}+\frac{4 f^2
	\left(y+15 y^3\right)}{\left(1+y^2\right)^3}\right) f'+\left(\frac{12}{1+y^2}\dvvtag+\frac{4 f
	\left(-15+4 y^2\right)}{\left(1+y^2\right)^2}\right) \left(f'\right)^2-\frac{18 y
	\left(f'\right)^3}{1+y^2}+\left(\frac{f^2 \left(48-44
	y^2\right)}{\left(1+y^2\right)^2}-\frac{8 f}{1+y^2}+\frac{30 f y f'}{1+y^2}\right)
f''\Bigg]\, ,
\end{align}

\begin{align}\notag
C&=a \left(1-\frac{4 f}{1+y^2}\right)+\frac{\lambda}{a^3}  \left[\frac{20 f \left(-1+3
	y^2\right)}{\left(1+y^2\right)^3}+\frac{12 f^2 \left(9-50 y^2+5
	y^4\right)}{\left(1+y^2\right)^4}\dvv+\frac{64 f^3 \left(5-41 y^2+5
	y^4\right)}{\left(1+y^2\right)^5}+\left(\frac{48 f^2 y \left(25-13
	y^2\right)}{\left(1+y^2\right)^4}-\frac{40 y}{\left(1+y^2\right)^2}+\frac{f \left(732 y-148
	y^3\right)}{\left(1+y^2\right)^3}\right) f'\dvv
+\left(\frac{8 f \left(19+9
	y^2\right)}{\left(1+y^2\right)^3}+\frac{2 \left(-83+41
	y^2\right)}{\left(1+y^2\right)^2}\right) \left(f'\right)^2+\frac{88 y
	\left(f'\right)^3}{\left(1+y^2\right)^2}+\left(\frac{10}{1+y^2}\dvvv
+\frac{192 f^2 \left(-1+2
	y^2\right)}{\left(1+y^2\right)^3}+\frac{f \left(-76+64
	y^2\right)}{\left(1+y^2\right)^2}+\left(-\frac{184 f y}{\left(1+y^2\right)^2}-\frac{66
	y}{1+y^2}\right) f'-\frac{20 \left(f'\right)^2}{1+y^2}\right) f''\dvv
+\left(5+\frac{40
	f}{1+y^2}\right) \left(f''\right)^2+\left(-\frac{80 f^2 y}{\left(1+y^2\right)^2}-\frac{10 f
	y}{1+y^2}+\left(5+\frac{40 f}{1+y^2}\right) f'\right) f^{(3)}\right]\dv 
+\frac{\tilde\lambda}{a^3}
\left[\frac{48 f y}{\left(1+y^2\right)^3}+\frac{96 f^2 y \left(-7+4
	y^2\right)}{\left(1+y^2\right)^4}+\frac{224 f^3 y \left(-1+9
	y^2\right)}{\left(1+y^2\right)^5}+\left(\frac{6
	\left(-3+y^2\right)}{\left(1+y^2\right)^2}\dvvv -\frac{32 f^2 \left(4+55
	y^2\right)}{\left(1+y^2\right)^4}
+\frac{6 f \left(59-108
	y^2+y^4\right)}{\left(1+y^2\right)^3}\right) f'+\left(-\frac{6 y
	\left(-39+y^2\right)}{\left(1+y^2\right)^2}\dvvv +\frac{16 f y
	\left(4+y^2\right)}{\left(1+y^2\right)^3}\right) \left(f'\right)^2+\frac{\left(84-20
	y^2\right) \left(f'\right)^3}{\left(1+y^2\right)^2}+\left(-\frac{32 f^2 y
	\left(-21+y^2\right)}{\left(1+y^2\right)^3}-\frac{6 y}{1+y^2}\dvvv -\frac{2 f y \left(-71+3
	y^2\right)}{\left(1+y^2\right)^2}+\left(\frac{8 f \left(-24+5
	y^2\right)}{\left(1+y^2\right)^2}+\frac{-77+9 y^2}{1+y^2}\right) f'+\frac{12 y
	\left(f'\right)^2}{1+y^2}\right) f''\dvv
+\left(-3 y-\frac{24 f y}{1+y^2}\right)
\left(f''\right)^2+\left(2+\frac{16 f^2 \left(-6+y^2\right)}{\left(1+y^2\right)^2}+\frac{2 f
	\left(2+y^2\right)}{1+y^2}\dvvvtag
+\left(-3 y-\frac{24 f y}{1+y^2}\right) f'\right) f^{(3)}\right]\, ,
\end{align}

\begin{align}\notag
D&=-1+\frac{f \left(1-y^2\right)}{1+y^2}-2 y f'+\frac{1}{2}
\left(-1-y^2\right) f''+\left(-1-y^2\right) \Lambda  a^2+\frac{\lambda}{a^4}  \left[\frac{f^2 \left(8-40
	y^2\right)}{\left(1+y^2\right)^4}\dvv
-\frac{40 f^3 \left(1-10
	y^2+y^4\right)}{\left(1+y^2\right)^5}+\left(\frac{48 f^2 y \left(9-2
	y^2\right)}{\left(1+y^2\right)^4}-\frac{88 f y}{\left(1+y^2\right)^3}\right)
f'+\left(\frac{50-6 y^2}{\left(1+y^2\right)^2}\dvvv
+\frac{f \left(-298+312 y^2-6
	y^4\right)}{\left(1+y^2\right)^3}\right) \left(f'\right)^2+\frac{2 y \left(-55+3 y^2\right)
	\left(f'\right)^3}{\left(1+y^2\right)^2}+\left(\frac{8 f}{\left(1+y^2\right)^2}\dvvv
-\frac{12 f^2
	\left(-1+9 y^2\right)}{\left(1+y^2\right)^3}+\left(\frac{12 y}{1+y^2}+\frac{4 f y \left(-4+3
	y^2\right)}{\left(1+y^2\right)^2}\right) f'+\frac{\left(25-3 y^2\right)
	\left(f'\right)^2}{1+y^2}\right) f''\dvv
-\frac{20 f \left(f''\right)^2}{1+y^2}+\left(\frac{40
	f^2 y}{\left(1+y^2\right)^2}-\frac{20 f f'}{1+y^2}\right) f^{(3)}\right]+\frac{\tilde\lambda}{a^4}
\left[\frac{144 f^2 y}{\left(1+y^2\right)^4}-\frac{16 f^3 y \left(35+33
	y^2\right)}{\left(1+y^2\right)^5}\dvv
+\left(\frac{24 f \left(-4+3
	y^2\right)}{\left(1+y^2\right)^3}+\frac{8 f^2 \left(50-19 y^2+9
	y^4\right)}{\left(1+y^2\right)^4}\right) f'+\left(\frac{8 f y \left(56-13
	y^2\right)}{\left(1+y^2\right)^3}-\frac{48 y}{\left(1+y^2\right)^2}\right)
\left(f'\right)^2\dvv
+\frac{\left(-84+40 y^2\right)
	\left(f'\right)^3}{\left(1+y^2\right)^2}+\left(-\frac{8 f y}{\left(1+y^2\right)^2}-\frac{8
	f^2 y \left(24+y^2\right)}{\left(1+y^2\right)^3}+\left(\frac{16}{1+y^2}+\frac{4 f \left(6+7
	y^2\right)}{\left(1+y^2\right)^2}\right) f'\dvvvtag
-\frac{24 y \left(f'\right)^2}{1+y^2}\right)
f''+\frac{12 f y \left(f''\right)^2}{1+y^2}+\left(-\frac{8 f^2
	\left(-6+y^2\right)}{\left(1+y^2\right)^2}-\frac{8 f}{1+y^2}+\frac{12 f y f'}{1+y^2}\right)
f^{(3)}\right]\, ,
\end{align}

\begin{align}\notag
E&=-a f y+\frac{\lambda}{a^3}  \left[-\frac{32 f^2 y}{\left(1+y^2\right)^3}+\frac{f^3 \left(544 y-704
	y^3\right)}{\left(1+y^2\right)^4}+\left(-\frac{2 f \left(1+9
	y^2\right)}{\left(1+y^2\right)^2}\dvvv
-\frac{2 f^2 \left(91-492 y^2+9
	y^4\right)}{\left(1+y^2\right)^3}\right) f'+\frac{2 f y \left(-131+9 y^2\right)
	\left(f'\right)^2}{\left(1+y^2\right)^2}+\left(\frac{18 f y}{1+y^2}+\frac{2 f^2 y
	\left(-125+9 y^2\right)}{\left(1+y^2\right)^2}\dvvv
+\frac{f \left(83-15 y^2\right)
	f'}{1+y^2}\right) f''+3 f y \left(f''\right)^2+\left(-6 f-\frac{6 f^2
	\left(-4+y^2\right)}{1+y^2}+3 f y f'\right) f^{(3)}\right]\dv +\frac{\tilde\lambda}{a^3}
\left[\frac{12 f^2 \left(-1+15 y^2\right)}{\left(1+y^2\right)^3}+\frac{4 f^3 \left(29-342
	y^2+45 y^4\right)}{\left(1+y^2\right)^4}+\left(-\frac{96 f y}{\left(1+y^2\right)^2}\dvvv
-\frac{20
	f^2 y \left(-49+15 y^2\right)}{\left(1+y^2\right)^3}\right) f'+\frac{2 f \left(-51+61
	y^2\right) \left(f'\right)^2}{\left(1+y^2\right)^2}+\left(\frac{14 f}{1+y^2}+\frac{4 f^2
	\left(-27+26 y^2\right)}{\left(1+y^2\right)^2}\dvvvtag
-\frac{78 f y f'}{1+y^2}\right) f''+9 f
\left(f''\right)^2+\left(-\frac{18 f^2 y}{1+y^2}+9 f f'\right) f^{(3)}\right]\, ,
\end{align}

\begin{align}\notag
F&=-\frac{f \left(3+y^2\right)}{1+y^2}-2 y f'-\frac{1}{2}
\left(1+y^2\right) f''-\left(1+y^2\right) \Lambda a^2 +\frac{\lambda}{a^4}  \left[\frac{32 f^2 \left(-1+5
	y^2\right)}{\left(1+y^2\right)^4}\dvv
+\frac{8 f^3 \left(35-278 y^2+35
	y^4\right)}{\left(1+y^2\right)^5}+\left(\frac{48 f^2 y \left(34-15
	y^2\right)}{\left(1+y^2\right)^4}-\frac{56 f y}{\left(1+y^2\right)^3}\right)
f'-\left(\frac{20}{\left(1+y^2\right)^2}\dvvv
+\frac{2 f \left(73-192 y^2+3
	y^4\right)}{\left(1+y^2\right)^3}\right) \left(f'\right)^2+\frac{2 y \left(-11+3 y^2\right)
	\left(f'\right)^3}{\left(1+y^2\right)^2}+\left(\frac{16 f}{\left(1+y^2\right)^2}\dvvv
+\frac{12
	f^2 \left(-15+23 y^2\right)}{\left(1+y^2\right)^3}+\frac{4 f y \left(-50+3 y^2\right)
	f'}{\left(1+y^2\right)^2}+\frac{\left(5-3 y^2\right) \left(f'\right)^2}{1+y^2}\right)
f''+\frac{20 f \left(f''\right)^2}{1+y^2}\dvv
+\left(-\frac{40 f^2
	y}{\left(1+y^2\right)^2}+\frac{20 f f'}{1+y^2}\right) f^{(3)}\right]+\frac{\tilde\lambda}{a^4}
\left[\frac{16 f^3 y \left(-49+93 y^2\right)}{\left(1+y^2\right)^5}+\left(\frac{24
	f}{\left(1+y^2\right)^3}\dvvv
+\frac{8 f^2 \left(34-239 y^2+9
	y^4\right)}{\left(1+y^2\right)^4}\right) f'+\left(\frac{12 y}{\left(1+y^2\right)^2}-\frac{8
	f y \left(-64+11 y^2\right)}{\left(1+y^2\right)^3}\right) \left(f'\right)^2+\frac{20 y^2
	\left(f'\right)^3}{\left(1+y^2\right)^2}\dvv
+\left(-\frac{40 f^2 y
	\left(-12+y^2\right)}{\left(1+y^2\right)^3}-\frac{16 f
	y}{\left(1+y^2\right)^2}+\left(-\frac{4}{1+y^2}+\frac{4 f \left(-42+17
	y^2\right)}{\left(1+y^2\right)^2}\right) f'-\frac{12 y \left(f'\right)^2}{1+y^2}\right)
f''\dvvtag
-\frac{12 f y \left(f''\right)^2}{1+y^2}+\left(\frac{8 f^2
	\left(-6+y^2\right)}{\left(1+y^2\right)^2}+\frac{8 f}{1+y^2}-\frac{12 f y f'}{1+y^2}\right)
f^{(3)}\right]\, ,
\label{Ffunc}
\end{align}

\begin{align}\notag
G&=\frac{f}{a \left(1+y^2\right)}+\frac{\lambda}{a^5}  \left[\frac{f^2 \left(16-80
		y^2\right)}{\left(1+y^2\right)^4}-\frac{16 f^3 \left(5-41 y^2+5
		y^4\right)}{\left(1+y^2\right)^5}+\left(\frac{28 f y}{\left(1+y^2\right)^3}\dvvv
	+\frac{12 f^2 y
		\left(-25+13 y^2\right)}{\left(1+y^2\right)^4}\right)
	f'+\left(\frac{10}{\left(1+y^2\right)^2}-\frac{2 f \left(19+9
		y^2\right)}{\left(1+y^2\right)^3}\right) \left(f'\right)^2-\frac{22 y
		\left(f'\right)^3}{\left(1+y^2\right)^2}\dvv
	+\left(\frac{f^2 \left(48-96
		y^2\right)}{\left(1+y^2\right)^3}-\frac{8 f}{\left(1+y^2\right)^2}+\frac{46 f y
		f'}{\left(1+y^2\right)^2}+\frac{5 \left(f'\right)^2}{1+y^2}\right) f''-\frac{10 f
		\left(f''\right)^2}{1+y^2}+\left(\frac{20 f^2 y}{\left(1+y^2\right)^2}\dvvv
	-\frac{10 f
		f'}{1+y^2}\right) f^{(3)}\right]+\frac{\tilde\lambda}{a^5} \left[\frac{56 f^3 y \left(1-9
		y^2\right)}{\left(1+y^2\right)^5}+\left(-\frac{12 f}{\left(1+y^2\right)^3}+\frac{8 f^2
		\left(4+55 y^2\right)}{\left(1+y^2\right)^4}\right) f'\dvv
	-\left(\frac{6
		y}{\left(1+y^2\right)^2}+\frac{4 f y \left(4+y^2\right)}{\left(1+y^2\right)^3}\right)
	\left(f'\right)^2+\frac{\left(-21+5 y^2\right)
		\left(f'\right)^3}{\left(1+y^2\right)^2}+\left(\frac{8 f^2 y
		\left(-21+y^2\right)}{\left(1+y^2\right)^3}\dvvv
	+\frac{8 f y}{\left(1+y^2\right)^2}+\left(\frac{f
		\left(48-10 y^2\right)}{\left(1+y^2\right)^2}+\frac{2}{1+y^2}\right) f'-\frac{3 y
		\left(f'\right)^2}{1+y^2}\right) f''+\frac{6 f y \left(f''\right)^2}{1+y^2}\dvvtag
	+\left(-\frac{4
		f^2 \left(-6+y^2\right)}{\left(1+y^2\right)^2}-\frac{4 f}{1+y^2}+\frac{6 f y
		f'}{1+y^2}\right) f^{(3)}\right]\, .
	\label{Gfunc}
\end{align}

By direct computation, one can check that $\diff\boldsymbol{Q}_{\xi}=0$ once the equation of motion \req{feqcubicn} is assumed. 

\subsection{Proof that the $\xi^0$-integral vanishes}\label{app:rintegral}

Let us now show that the charge associated to the timelike Killing vector $\partial_{t}$ vanishes. To this end, we need to evaluate the integral proportional to $\xi^0$ in \req{Jint1}, this is

\begin{equation}
\int_{x_1}^{x_2}dx F(x)\, .
\end{equation}
It is possible to check that

\begin{equation}
F=-\frac{a^2}{a^2+x^2} \frac{d \E_{f}}{d x}-\frac{1}{2} x \frac{d^2\E_{f}}{d x^2}+\frac{d\mathcal{F} }{d x}\, ,
\end{equation}
where $\E_{f}$ is the equation of motion for $f$, given by \req{feqcubic}, and 

\begin{align}\notag
\mathcal{F}&=f\Bigg\{-x +\lambda   \left[-\frac{32 a^2 f x}{\left(a^2+x^2\right)^3}+\frac{32 f^2 \left(17 a^4 x-22 a^2
	x^3\right)}{\left(a^2+x^2\right)^4}+\left(-\frac{2 \left(a^2+9
	x^2\right)}{\left(a^2+x^2\right)^2}\dvvv
-\frac{2 f \left(91 a^4-492 a^2 x^2+9
	x^4\right)}{\left(a^2+x^2\right)^3}\right) f'+\frac{\left(-262 a^2 x+18 x^3\right)
	\left(f'\right)^2}{\left(a^2+x^2\right)^2}+\left(\frac{18 x}{a^2+x^2}\dvvv
+\frac{f \left(-250 a^2
	x+18 x^3\right)}{\left(a^2+x^2\right)^2}+\frac{\left(83 a^2-15 x^2\right)
	f'}{a^2+x^2}\right) f''+3 x \left(f''\right)^2+\left(-6+\frac{f \left(24 a^2-6
	x^2\right)}{a^2+x^2}\dvvv+3 x f'\right) f^{(3)}\right]+\tilde\lambda \left[-\frac{12 f
	\left(a^3-15 a x^2\right)}{\left(a^2+x^2\right)^3}+\frac{4 f^2 \left(29 a^5-342 a^3 x^2+45 a
	x^4\right)}{\left(a^2+x^2\right)^4}+\left(-\frac{96 a x}{\left(a^2+x^2\right)^2}\dvvv
+\frac{20 f
	\left(49 a^3 x-15 a x^3\right)}{\left(a^2+x^2\right)^3}\right) f'-\frac{2 \left(51 a^3-61 a
	x^2\right) \left(f'\right)^2}{\left(a^2+x^2\right)^2}+\left(\frac{14 a}{a^2+x^2}\dvvv
-\frac{4 f
	\left(27 a^3-26 a x^2\right)}{\left(a^2+x^2\right)^2}-\frac{78 a x f'}{a^2+x^2}\right) f''+9
a \left(f''\right)^2+\left(-\frac{18 a f x}{a^2+x^2}+9 a f'\right) f^{(3)}\right]\Bigg\}\, .
\end{align}
Then, on-shell $\E_{f}=n$ is constant, and therefore we have

\begin{equation}
\int_{x_1}^{x_2}dx F(x)=\int_{x_1}^{x_2}dx \frac{d\mathcal{F} }{d x}=\mathcal{F}(x)\Big|^{x_2}_{x_1}=0\, ,
\end{equation}
which vanishes because $\mathcal{F}$ is proportional to $f$, which vanishes at the poles $x=x_1,x_2$. Thus, the Komar charge is indeed independent of the radius.

\bibliographystyle{JHEP}
\bibliography{Gravities.bib}

\end{document}